\pdfoutput=1
\documentclass[aps,showpacs,twocolumn,twoside,pra,superscriptaddress]{revtex4-2}
\usepackage{epsfig,amsmath,amssymb,amsfonts}
\usepackage{graphicx}
\usepackage{array}
\usepackage{xcolor}
\newcommand{\be}{\begin{equation}}
\newcommand{\ee}{\end{equation}}

\newcommand{\ra}{\rangle}

\def\|#1{\!#1\!}
\def\vecb#1{\boldsymbol{#1}}

\def\matr#1#2#3{\langle#1|#2|#3\rangle}

\def\E#1{\cdot10^{#1}}

\def\={\!=\!}
\def\>{\!>\!}
\def\<{\!<\!}
\def\-{\!-\!}
\def\+{\!+\!}

\def\R#1{#1_{\rm R}}
\def\I#1{#1_{\rm I}}
\def\uvo#1{\lq\lq #1\rq\rq}

\begin{document}

\title{Continuum analogues of excited-state quantum phase transitions}

\author{Pavel Str\'ansk\'y}
\email{stransky@ipnp.mff.cuni.cz}
\affiliation{Institute of Particle and Nuclear Physics, Faculty of Mathematics and Physics, Charles University, V Hole\v{s}ovi\v{c}k\'ach 2, 18000 Prague, Czechia} 
\author{Milan {\v S}indelka}
\email{sindelka@ipp.cas.cz}
\affiliation{Institute of Plasma Physics, Academy of Sciences of the Czech Republic, Za Slovankou 3, 18200 Prague, Czechia} 
\author{Pavel Cejnar}
\email{cejnar@ipnp.mff.cuni.cz}
\affiliation{Institute of Particle and Nuclear Physics, Faculty of Mathematics and Physics, Charles University, V Hole\v{s}ovi\v{c}k\'ach 2, 18000 Prague, Czechia}

\date{\today}

\begin{abstract}
Following our work [Phys.\,Rev.\,Lett.\,125,\,020401 (2020)], we discuss a semiclassical description of one-dimensional quantum tunneling through multibarrier potentials in terms of complex time. 
We start by defining a complex-extended continuum level density of unbound systems and show its relation to a complex time shift of the transmitted wave. 
While the real part of the level density and time shift describes the passage of the particle through classically allowed coordinate regions, the imaginary part is connected with an instanton-like picture of the tunneling through forbidden regions. 
We describe singularities in the real and imaginary parts of the level density and time shift caused by stationary points of the tunneling potential, and show that they represent a dual extension of excited-state quantum phase transitions from bound to continuum systems.
Using the complex scaling method, we numerically verify the predicted effects in several tunneling potentials.
\end{abstract}

\maketitle

\section{Introduction}

During the past 20 years, the research of interacting quantum systems has been enforced on both experimental and theoretical sides in response to growing possibilities to prepare, probe and utilize customized laboratory quantum systems \cite{Gar14+}.
An important direction of this research concerns new manifestations of various critical phenomena \cite{Car10}.
While the thermal and quantum phase transitions affect static properties of systems in equilibrated states, novel types of criticality apply also to the dynamical properties associated with non-thermal excitations.
In particular, the so-called dynamical quantum phase transitions denote non-analyticities in the evolution of an initially equilibrated state after a sudden change of a control parameter \cite{Hey18}.
The excited-state quantum phase transitions (ESQPTs), on the other hand, represent sharp changes observed directly in the spectra of excited states---both in the arrangement of energy eigenvalues (the density and slope of energy levels in the {energy\,$\times$\,parameter} plane) and in the form of energy eigenvectors.
These features encode the dynamics of the system in the given excitation domain. 
Numerous examples of ESQPTs can be found, e.g., in Refs.\,\cite{Cej06,Cap08,Bra13,Lar13,Die13,Str14,Bas14a,Bas14,Rel14,Kop15,Pue16,Str16,Sin17,Byc18,Gar18,Kha19,Mac19,Pue20,Tia20,Fel20} and a recent review in Ref.\,\cite{Cej20}.

The above-mentioned types of quantum criticality are usually investigated in bound systems, i.e., systems whose Hamiltonians yield
localized (normalizable) eigenstates and discrete spectra of energy levels. 
However, our recent work in Ref.\,\cite{Str20} demonstrated that close analogues of ESQPTs exist also in unbound quantum systems with continuous energy spectra and unnormalizable eigenstates, namely in one-dimensional (1D) tunneling systems with arbitrary potentials.
The quantity forming a counterpart of the discrete level density of bound systems is the so-called continuum level density, whose complex extension encodes full information on the tunneling (transmission) amplitudes.
We showed that the continuum level density exhibits singularities connected with stationary points of the tunneling potential. 
The singularities appear in both real and imaginary parts of the complex-extended continuum level density, generalizing the corresponding types of ESQPTs of 1D bound systems to a dual form associated with the real and instanton-like segments of the tunneling trajectories \cite{Str20}.

Potential applications of these analyses may be extensive. 
Quantum tunneling processes attract no less of recent attention than the dynamics and spectroscopy of bound quantum systems \cite{Raz14}.
This is partly due to the fundamental importance of such processes in the nature and technology \cite{Col77a,Nobel}, but also because of rapid theoretical and experimental progress in several related areas.
Let us mention (giving only some example references) the investigations of tunneling in driven and open systems \cite{Gri98}, advances of dynamical tunneling \cite{Kes11}, the long-standing study of tunneling times \cite{Hen01,Sha12,Lan15,Sat19,Ram20}, and also practical realizations of customized tunneling potentials by means of suitable nanostructures \cite{See01,Bha06,Suz10,Bri13,Gol15,Tao19}.

\begin{figure*}[t]
\includegraphics[width=0.9\linewidth]{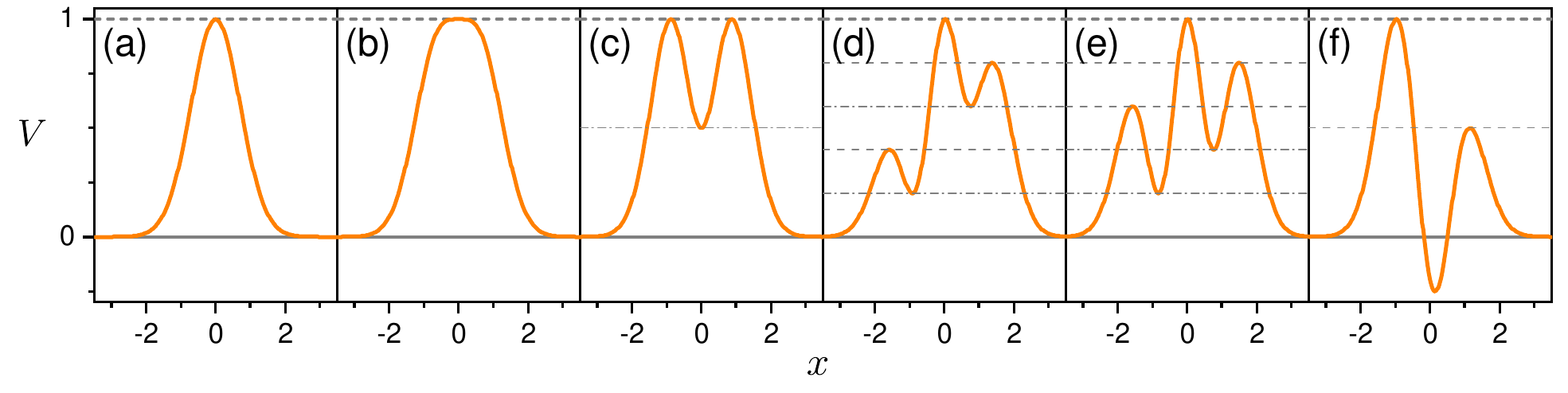}
\caption{
Sample potentials $V(x)$ from Eq.\,\eqref{Ham} with polynomials up to the quartic term. 
Parameters $(c_0,c_1,c_2,c_3,c_4)$ are: (a)  $(1,0,0,0,0)$, (b) $(1, 0, 1, 0, 0)$, (c) $(0.5, 0, 2.156, 0, 0)$, (d) $(1, -0.138, -1.278, -0.485, 1.473)$, (e) $(1, -0.072, -1.921, -0.260, 1.961)$, and (f) $(-0.197, -0.718, 2.192, 0)$; in all panels ${\eta=1}$.
Energies of stationary points (marked by the horizontal lines) from left to right in each panel are (a) 1, (b) 1, (c) 1, 0.5, 1, (d) 0.4, 0.2, 1, 0.6, 0.8, (e) 0.6, 0.2, 1, 0.4, 0.8, and (f) 1, $-0.25$, 0.5.
All stationary points except in panel (b) are quadratic, the one in panel (b) is quartic. 
These potentials serve as illustrative examples in the forthcoming analyses.
}
\label{pots}
\end{figure*}

The purpose of this paper is to elucidate and extend the results of of our initial analysis of ESQPT-like tunneling singularities from Ref.\,\cite{Str20}.
We consider a 1D tunneling problem with a single-particle Hamiltonian 
\begin{equation}
\hat{H}=\underbrace{\frac{\hat{p}^2}{2m}}_{\hat{H}^{(0)}}+\underbrace{\left(c_0+c_1x+c_2x^2+\dots\right)e^{-\eta x^2}}_{\hat{V}(x)}, 
\label{Ham}
\end{equation} 
where $\hat{H}^{(0)}$ stands for the free Hamiltonian (with $\hat{p}\|{\equiv}-i\hbar\frac{\partial}{\partial x}$ denoting the momentum operator and $m$ the particle mass) and ${\hat{V}(x)\|{\equiv}V(x)}$ for the interaction potential with adjustable parameters $c_0,c_1,c_2,...$ and $\eta$.
Below we set ${m=1}$, ${\eta=1}$ and use potentials with polynomials up to the quartic term.
Specific forms of these potentials used in this work are depicted in Fig.\,\ref{pots}.
Due to its Gaussian attenuation, the potential function $V(x)$ is negligible outside a certain finite interval $(a,b)$, where ${a<0}$ and ${b>0}$ are two points sufficiently far from the origin.
To solve the tunneling problem, we therefore require the standard asymptotics of wave functions, namely
\begin{equation}
\psi(x)=\left\{\begin{array}{ll}
e^{+ipx/\hbar}\+\alpha(E)e^{-ipx/\hbar} & {\rm for\ }x\<a,
\\
\beta(E)e^{+ipx/\hbar} & {\rm for\ }x\>b,
\end{array}\right.
\label{asy}
\end{equation} 
where $p\=\sqrt{2mE}$, and $\alpha(E)$ and $\beta(E)$, respectively, stand for the reflection and transmission amplitudes, satisfying the normalization condition ${|\alpha(E)|^2+|\beta(E)|^2=1}$.
The transmission amplitude is written as
\begin{equation}
\beta(E)\=
e^{i\Phi(E)}
\label{tra},
\end{equation}
where ${\Phi(E)\in{\mathbb C}}$ is a complex phase that encodes both observable quantities associated with the 1D tunneling process, namely the transmission probability $|\beta(E)|^2$ and the real phase shift $\varphi(E)$ of the transmitted wave:
\begin{equation}
|\beta(E)|^2=e^{-2\,{\rm Im}\Phi(E)},\quad
\varphi(E)={\rm Re}\,\Phi(E).
\label{cop}
\end{equation}

The plan of the paper is the following.
In Sec.\,\ref{Leden} we introduce a complex density of continuum states associated with Hamiltonians of the general form ${\hat{H}=\hat{H}^{(0)}+\hat{V}(x)}$ with finite-range potentials and show that it fully describes the complex transmission amplitude in Eq.\,\eqref{tra}.
In Sec.\,\ref{Cosca} we explain the determination of the complex continuum level density with the aid of the complex scaling method, showing illustrative numerical examples with potentials from Fig.\,\ref{pots}.
In Sec.\,\ref{Times} we overview the connection of the continuum level density with the semiclassical time shift of the transmitted wave and derive its complex extension in terms of instanton-like tunneling trajectories.
In Sec.\,\ref{Esqpts} we present a typology of ESQPT-like singularities of the continuum level density and time shifts connected with stationary points of the tunneling potentials.
The theoretical results are again illustrated by numerical examples based on sample potentials from Fig.\,\ref{pots}.
Section~\ref{Clouseau} gives a brief summary and conclusion.

\section{Complex extension of the continuum level density}
\label{Leden}

\subsection{Continuum level density}
\label{Contle}

The level density at energy $E$ for discrete energy spectra of bound quantum systems is defined as
\begin{eqnarray}
\rho(E)&=&\sum_{k}\delta(E-E_k)=\lim_{\epsilon\to 0+}\sum_k\underbrace{\frac{1}{\pi}\frac{\epsilon}{(E\-E_k)^2+\epsilon^2}}_{C_{2\epsilon}(E\-E_k)}
\nonumber\\
&=&-\frac{1}{\pi}\lim_{\epsilon\to 0+}{\rm Im}\,{\rm Tr}\,\hat{G}(E\+i\epsilon)
\label{led},
\end{eqnarray}
where $E_k$ (with $k\=1,2,...$) are discrete eigenvalues of the Hamiltonian $\hat{H}$ and 
\begin{equation}
\hat{G}({\cal E})=\frac{1}{{\cal E}-\hat{H}}
\label{Gr}
\end{equation} 
is the Green operator at complex energy ${{\cal E}=E\+i\epsilon}$.
Note that here, the infinitesimal imaginary part $\epsilon$ is added to the real energy $E$ to prevent the divergences of $\hat{G}(E)$ at $E\=E_k$.
Hence each $\delta$-function in the first sum is turned into the normalized Cauchy (Breit-Wigner) peak $C_{\Gamma}(E\-E_0)$ with maximum $E_0$ located at the respective level energy $E_k$ and the full width at half maximum $\Gamma$ equal to $2\epsilon$.

The level density has been introduced also for unbound systems with continuous energy spectra \cite{Lev69,Kru98,Kru99}.
Consider such a system with Hamiltonian $\hat{H}=\hat{H}^{(0)}\+\hat{V}$, including the free term $\hat{H}^{(0)}$ and an interaction term $\hat{V}$, and its Green operator \eqref{Gr}. 
The continuum level density is defined as
\begin{equation}
\delta\rho(E)=-\frac{1}{\pi}\lim_{\epsilon\to 0+}{\rm Im}\,{\rm Tr}\left[\hat{G}(E\+i\epsilon)\-\hat{G}^{(0)}(E\+i\epsilon)\right]
\label{olde},
\end{equation}
where 
\begin{equation}
\hat{G}^{(0)}({\cal E})=\frac{1}{{\cal E}-\hat{H}^{(0)}}
\label{Gr0}
\end{equation}
is the Green operator of the free Hamiltonian.
The trace in Eq.\,\eqref{olde} is in principle performed by integration of the expectation value of  ${[\hat{G}(E\+i\epsilon)-\hat{G}^{(0)}(E\+i\epsilon)]}$ in a continuous basis of stationary scattering states.
A more intuitive approach is to calculate the trace in the discrete energy basis associated with the system confined to a sufficiently large but finite coordinate interval of length $L$ containing the whole spatial support of the interaction $\hat{V}$ (or to a finite box for higher than 1D problems).
The trace in Eq.\,\eqref{olde} is then obtained as the ${L\to\infty}$ limit of the corresponding finite-size expressions.
This indicates that for ${L\to\infty}$, separate traces of both Green operators are infinite and only the trace of their difference yields a well defined density $\delta\rho(E)$.

Because the quantity \eqref{olde} can obviously be negative (in the finite-$L$ approximation it is a difference of two semi-positive densities associated with the two Green operators), it prevents an interpretation as a kind of continuous weight function. 
However, as proven in Ref.\,\cite{Lev69}, the energy dependence of $\delta\rho(E)$ contains complete information on the behavior of the real phase shift of the transmitted wave.
In particular, for 1D problems with asymptotic wave functions of the form from Eqs.\,\eqref{asy} and \eqref{tra}, the continuum level density \eqref{olde} is connected with the real phase shift $\varphi(E)$ by the following relation:
\begin{equation}
\delta\rho(E)=\frac{1}{\pi}\frac{d}{dE}\varphi(E)
\label{dphi}.
\end{equation}
This means that $\varphi(E)$ can be uniquely (up to an arbitrary additional constant) determined by integration of $\delta\rho(E)$.
The formula \eqref{dphi} is (in a more general form) derived in Appendix \ref{AppA}.

\subsection{Complex continuum level density}
\label{Cocontle}

\begin{figure}[t]
\includegraphics[width=\linewidth]{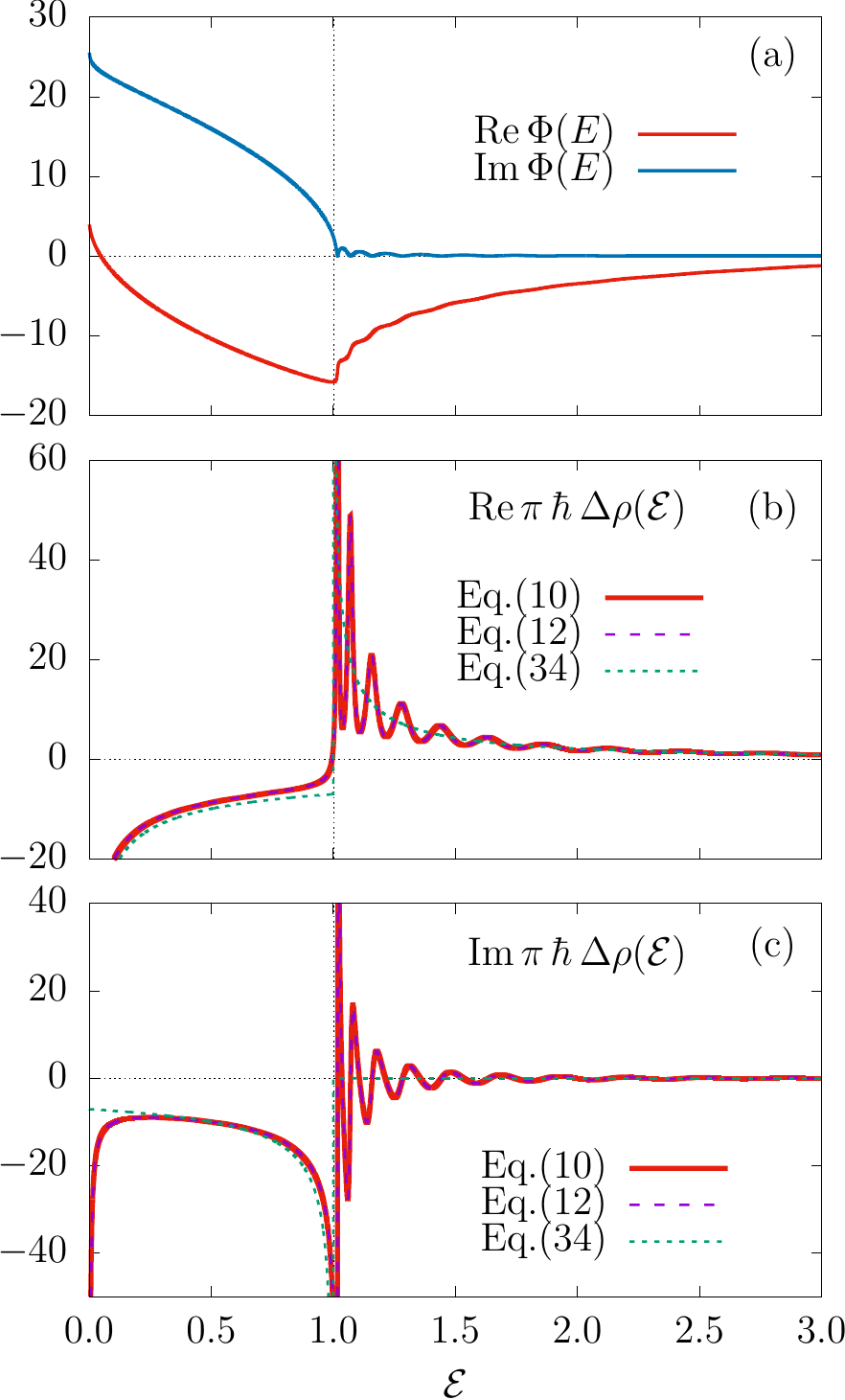}
\caption{
The complex phase $\Phi(E)$ and the complex continuum level density $\Delta\rho({\cal E})$ for a rectangular potential barrier $V(x)$, set to ${V=1}$ for ${|x|\leq 5}$ and ${V=0}$ for ${|x|>5}$, calculated with ${m=1}$ and ${\hbar=0.6}$.
Panel (a) shows the real and imaginary parts of $\Phi(E)$ from an analytic expression of the complex transmission amplitude \eqref{tra}.
Panels (b) and (c), respectively, display the real and imaginary parts of $\Delta\rho({\cal E})$ calculated from the defining formula \eqref{lev} (the full curve) as well as from Eq.\,\eqref{dPhi} (the dashed curve). 
The semiclassically smoothed densities ${\rm Re}\,\Delta{\overline\rho}({\cal E})$ and ${\rm Im}\,\Delta{\overline\rho}({\cal E})$ (dotted curves) are determined from Eq.\,\eqref{deti}. 
}
\label{recta}
\end{figure}

In this paper, following Ref.\,\cite{Str20}, we introduce a com\-plex-extended continuum level density by the formula 
\begin{equation}
\Delta\rho({\cal E})=\frac{i}{\pi}\,{\rm Tr}\left[\hat{G}({\cal E})-\hat{G}^{(0)}({\cal E})\right]
\label{lev}.
\end{equation}
It coincides with the formula \eqref{olde} without the Im symbol and with the complex energy ${\cal E}$ not restricted to $E\+i\epsilon$.
Note also that in the definition \eqref{lev} we use the imaginary factor $i$ to make the ${\rm Im}\leftrightarrow{\rm Re}$ conversion, which turns out convenient in the forthcoming considerations.

The density $\Delta\rho({\cal E})$ takes complex values and is defined in the complex energy plane ${\cal E}\=E\-\frac{i}{2}\Gamma$.
The imaginary part of ${\cal E}$ is by default taken negative since some discrete, isolated states with ${\rm Im}\,{\cal E}\<0$ can be interpreted as resonances.
This approach will be elaborated in Sec.\,\ref{Cosca}.
However, we will be mostly interested in the behavior of Eq.\,\eqref{lev} on the real axis, i.e., at ${\cal E}=E\-i0=E$.
Here we can write
\begin{equation}
{\rm Re}\,\Delta\rho(E)
=\delta\rho(E),
\label{tauto}
\end{equation}
which means that for 1D scattering problems, the real density ${\rm Re}\,\Delta\rho(E)$ determines, through Eq.\,\eqref{dphi}, the real phase shift ${\rm Re}\,\Phi(E)=\varphi(E)$ from Eq.\,\eqref{tra}.
The imaginary density ${\rm Im}\,\Delta\rho(E)$ can be heuristically anticipated to do the same job for the imaginary phase shift ${\rm Im}\,\Phi(E)$.
Hence we extend Eq.\,\eqref{dphi} to a more general form 
\begin{equation}
\Delta\rho(E)=\frac{1}{\pi}\frac{d}{dE}\Phi(E)
\label{dPhi}.
\end{equation}
This means that the real part ${\rm Re}\,\Delta\rho(E)$ is given by formula \eqref{dphi}, while for the imaginary part we get
\begin{equation}
{\rm Im}\,\Delta\rho(E)=-\frac{1}{2\pi}\frac{d}{dE}\ln|\beta(E)|^2
\label{imdPhi}.
\end{equation}
A proof of Eq.\,\eqref{dPhi} for 1D systems is given in Appendix~\ref{AppA}.
An illustrative example is presented in Fig.\,\ref{recta}, where the formula \eqref{dPhi} is applied to the familiar rectangular potential barrier with analytically calculable transmission amplitude $\beta(E)$.
We see that the complex continuum level density evaluated from the Green operators via Eq.\,\eqref{lev} agrees perfectly with that calculated from the transmission coefficient via Eq.\,\eqref{dPhi}.

Formula \eqref{dPhi} implies that the transmission probability $|\beta(E)|^2$ and the phase of the transmitted wave $\varphi(E)$ from Eq.\,\eqref{cop} can be obtained through the integration of functions ${\rm Re}\Delta\rho(E)$ and ${\rm Im}\Delta\rho(E)$.
So the complex continuum level density $\Delta\rho(E)$ contains complete information on the transmission amplitude and, in principle, is accessible to experimental study.

\subsection{Smoothed level density and infinite-size limit}
\label{Smoole}

In finite quantum systems with discrete spectra, the level density $\rho(E)$ needs to be purged of finite-size oscillatory structures to extract its principal behavior.
The smoothed level density $\overline{\rho}(E)$ can be determined in two ways.
The first one is based on evaluating the sum in the second equality of Eq.\,\eqref{led} with a certain positive value of imaginary energy $\epsilon$, so $\overline{\rho}(E)={\rm Im}\,{\rm Tr}\,\hat{G}(E\+i\epsilon)$.
For $\epsilon$ exceeding a typical distance of energy levels, this procedure converts the chain of $\delta$-functions into a sum of mutually overlapping Cauchy functions (alternatively, one can also use the Gaussian or other smoothening functions), which yields the desired smooth dependence.

The second smoothening method is based on the semiclassical approximation.  
In particular, for a system with ${f=1}$ degree of freedom we can write \cite{Cap08,Str14}
\begin{equation}
\overline{\rho}(E)=\frac{1}{2\pi\hbar}\ \frac{d}{dE}\!\!\!\iint\limits_{H(q,p)\leq E}\!\!\!dq\,dp
=\frac{\tau(E)}{2\pi\hbar}
\label{Tper},
\end{equation}
where  the integral measures the volume of the space available for a system with classical Hamiltonian $H(q,p)$ at energies less than or equal to~$E$.
The derivative of the phase-space volume function for an ${f=1}$ system can be expressed via a sum of periods $\tau
(E)={\sum_k\tau_k(E)}$ of all primitive orbits at energy~$E$.
In systems with any finite value of~$f$, which does not increase with the system's size, the semiclassical limit coincides with the infinite-size limit and both the above-mentioned smoothening methods become equivalent \cite{Str14,Cej20}.

While \uvo{classicality} in the above considerations is measured by the value of the Planck constant $\hbar$, the size of a~many-body system is naturally defined by the number~$N$ of elementary constituents.
But how to define the size of a one-body system?
We consider an ${f=1}$ quantum Hamiltonian of the form~\eqref{Ham} with an arbitrary potential $V(x)$ and assume that both the coordinate $x$ and energy $E$ are dimensionless, measured in units of their typical scales $x_0$ and $E_0$, respectively. 
The Hamiltonian in these units reads ${\hat{H}=-(\partial/\partial x)^2/(2\varkappa^2)+V(x)}$, where
\begin{equation}
\varkappa=\frac{x_0\sqrt{mE_0}}{\hbar}.
\label{size}
\end{equation}
This dimensionless parameter represents a typical action in units of $\hbar$, which at the given energy scale plays a similar role as the number $N$ for an interacting many-body system at the energy scale ${E_0\sim N\hbar\omega}$, where $\hbar\omega$ is a typical elementary excitation energy in the one-body term of the Hamiltonian \cite{Str14,Cej20}.
Hence $\varkappa$ can be seen as a suitable size parameter for one-body systems.
It is obvious that the limit ${\varkappa\to\infty}$ is equivalent to ${\hbar\to 0}$, and therefore implies the validity of the semiclassical approximation.
In this limit, the ESQPTs emerge as singularities of the scaled level density ${\overline{\rho}(E)/\varkappa\propto\hbar\,\overline{\rho}(E)}$ \cite{Cap08,Str14,Cej20}.

The smoothening procedure is applied also to the continuum level density $\Delta\,\rho({\cal E})$ on the real energy axis ${\cal E}={E-i0}$.
Even in absence of bound states ($\delta$-functions), the dependencies ${\rm Re}\,\Delta\rho(E)$ and ${\rm Im}\,\Delta\rho(E)$ often contain rather sharp resonance contributions (see Sec.\,\ref{Cosca}), and to extract robust features of the scattering process, these structures need to be smoothed out. 
We again use two methods.
The first one is based on the above-explained trick with a small imaginary shift of energy, i.e., on replacing the real energy $E$ in Eq.\,\eqref{lev} by ${{\cal E}=E+i\epsilon}$ with a small ${\epsilon>0}$.
The second method---the one based on the semiclassical approximation---will be described in Sec.\,\ref{Comti}, where we will derive an analogue of Eq.\,\eqref{Tper}.
The result of the semiclassical smoothening for the square barrier is shown in Fig.\,\ref{recta}.

We again assert that both smoothening methods become equivalent in the ${\varkappa\to\infty}$ limit.
The smoothed complex density $\Delta\overline{\rho}(E)$ is related, via an analog of Eq.\,\eqref{dPhi}, to the smoothed complex phase shift ${{\overline\Phi}(E)\equiv{\overline\varphi}(E)\-\frac{i}{2}\overline{\ln|\beta(E)|^2}}$, which involves smoothening of the energy dependencies of both the phase and intensity of the transmitted wave.

\section{Continuum level density from the complex scaling method}
\label{Cosca}

\subsection{Complex scaling method}
\label{Coscame}

The complex scaling method is an efficient way of calculating cross sections or transmission probabilities in scattering processes involving resonances.
The method was introduced in Refs.\,\cite{Bal71,Sim72+}, reviewed in Refs.\,\cite{Ho83,Moi98,Moi11}, and elaborated in connection with the real continuum level density in Refs.\,\cite{Suz05,Suz08}.
It makes use of a similarity transformation with a non-unitary operator 
\begin{equation}
{\hat{S}_{\vartheta}=e^{i\vartheta/2}e^{-\vartheta\,\hat{x}\hat{p}/\hbar}}
\label{Sim}, 
\end{equation}
where the angle ${\vartheta\in(0,\vartheta_{\rm max})}$ is a fixed parameter, with the limiting value $\vartheta_{\rm max}$ set for Hamiltonians of the form~\eqref{Ham} (with the asymptotic Gaussian shape of the potential) to $\frac{\pi}{4}$.
The transformation maps the coordinate and momentum operators $\hat{x}$ and $\hat{p}$ to 
\begin{equation}
\hat{S}_{\vartheta}\hat{x}\hat{S}_{\vartheta}^{-1}=e^{i\vartheta}\hat{x},\qquad 
\hat{S}_{\vartheta}\hat{p}\hat{S}_{\vartheta}^{-1}=e^{-i\vartheta}\hat{p},
\label{xp}
\end{equation}
and the original Hamiltonian $\hat{H}$ to an equivalent non-Hermitian image 
\begin{equation}
\hat{S}_{\vartheta}\hat{H}\hat{S}_{\vartheta}^{-1}\equiv\hat{{\cal H}}_{\vartheta}=
\underbrace{e^{-2i\vartheta}\frac{\hat{p}^2}{2m}}_{\hat{{\cal H}}_{\vartheta}^{(0)}}
+V(e^{+i\vartheta}\hat{x}).
\end{equation}

The complex scaling method is used to identify discrete resonant solutions of the scattering problem with the full Hamiltonian~$\hat{H}$.
These solutions correspond to poles of the scattering matrix at complex momenta ${p_k=|p_k|e^{-i\gamma_k}}$, where ${k=1,2,...}$ is an enumerating index and ${\gamma_k\in(0,\frac{\pi}{2})}$ denotes a phase of $p_k$ in the ${\mathbb C}$ plane \cite{Kuk89}.
The resonance wave function $\psi_k(x)$ satisfies the outgoing (Siegert type) boundary conditions, i.e., it behaves for $x\to\pm\infty$ as $\sim e^{\pm ip_k x/\hbar}$.
The transformed wave function, which in the asymptotic region reads
\begin{equation}
\hat{S}_{\vartheta}\psi_k(|x|\|{\to}\infty)=e^{i\cos(\vartheta-\gamma_k)|p_k||x|/\hbar}\ e^{-\sin(\vartheta-\gamma_k)|p_k||x|/\hbar},
\label{sqint}
\end{equation}
is square integrable if ${\gamma_k\in(0,\vartheta)}$ (in the non-Hermitian formalism, the scalar product is replaced by a so-called c-product in which the bra function is not complex conjugate \cite{Moi11}).
The function $\hat{S}_{\vartheta}\psi_k(x)$, despite its unusual form with persistent oscillations in the asymptotic region, represents a normalizable eigenstate of the transformed Hamiltonian $\hat{{\cal H}}_{\vartheta}$.
Its energy ${{\cal E}_k=|p_k|^2e^{-i2\gamma_k}/(2m)}$ is complex, having the form ${{\cal E}_k=E_k-\frac{i}{2}\Gamma_k}$, where ${E_k>0}$ and ${\Gamma_k>0}$ are interpreted, respectively, as the centroid energy and the energy width of a resonance corresponding to the original Hamiltonian $\hat{H}$.

The resonances which under the transformation \eqref{Sim} become square integrable eigenstates of $\hat{{\cal H}}_{\vartheta}$ lie in the angular segment of the ${{\cal E}\in{\mathbb C}}$ plane given by the constraint ${\frac{1}{2}\Gamma_k/E_k\in(0,\tan 2\vartheta)}$ following from the ${\gamma_k<\vartheta}$ condition.
The number of poles satisfying this condition, i.e., the number of resonances identified by the complex scaling method with a given angle $\vartheta$, is denoted as $N_{\vartheta}$, so the enumerating index in ${\cal E}_k$ runs within the range ${k=1,2,...,N_{\vartheta}}$.

On the other hand, the resonances corresponding to the poles $p_k$ with phases ${\gamma_k>\vartheta}$ remain unrecognized by the transformation \eqref{Sim}.
They appear along the ${E>0}$ segment of the line in the ${{\cal E}\in{\mathbb C}}$ plane defined by the condition ${\frac{1}{2}\Gamma/E=\tan 2\vartheta}$.
This segment (ray) is referred to as the rotated continuum because it also carries continuous eigenstates of $\hat{{\cal H}}_{\vartheta}$, i.e, images of non-resonant eigensolutions of the full Hamiltonian $\hat{H}$. 
Note that in the finite-$L$ approximation, the unrecognized resonances as well as non-resonant solutions form discrete sets of states with energies ${{\cal E}_l=E_l-\frac{i}{2}\Gamma_l}$, which are all located along the rotated continuum ray.
To distinguish them from the identified resonances, we use for them a special enumerating index ${l=N_{\vartheta}\+1,N_{\vartheta}\+2,...}$.
Applying the same procedure based on the transformation \eqref{Sim} to the free Hamiltonian $\hat{H}^{(0)}$, which has no resonances at all, one finds all eigensolutions of $\hat{{\cal H}}^{(0)}_{\vartheta}$ located only along the rotated continuum ray.
In the finite-$L$ approximation we denote them as ${\cal E}^{(0)}_l=E^{(0)}_l\-\frac{i}{2}\Gamma^{(0)}_l$ with ${l=1,2,...}$.

As follows from the above explanations, outcomes of the complex scaling method depend on the selected angle $\vartheta$ in the similarity transformation \eqref{Sim}.
Only the resonances with ${\gamma_k<\vartheta}$ are seen to \uvo{condense} below the rotated continuum ray in the $E\times\Gamma$ plane, while all the other states localize along this ray.
However, it turns out that the physical results attributed to complex energies ${\cal E}$ below the rotated continuum ray, and particularly to real energies ${{\cal E}=E-i0}$, are entirely independent of $\vartheta$ in the ${L\to\infty}$ limit.
The proof of this statement is based on extended completeness relations in terms of the eigenstates of ${\cal H}_{\vartheta}$ for any $\vartheta$ \cite{Suz05,Suz08}.
If considering a sequence of calculations with decreasing angle~$\vartheta$, the independence of results on $\vartheta$ requires tiny redistributions of states on the rotated continuum ray which compensate effects of the disappearing resonances.
This remains so even for the angle ${\vartheta=0}$ assigned to the trivial transformation ${\hat{S}_{\vartheta}=1}$, when all results are obtained from a finite-$L$ diagonalization of the original Hamiltonian $\hat{H}$ and the ${L\to\infty}$ limiting procedure.
We stress, however, that the complex scaling method provides us a valuable physical picture in which the observed effects are properly attributed to the most important resonant solutions.
This holds also in calculations of the complex continuum level density.

\subsection{Evaluation of complex continuum level density}
\label{Coscale}

As shown in Refs.\,\cite{Str20,Suz05,Suz08}, the complex scaling method is very well suited for the calculation of the real as well as complex continuum level density.
In the finite-$L$ approximation, the trace in Eq.\,\eqref{lev} can be evaluated separately for both Green operators as a sum over all discrete eigenstates of the respective non-Hermitian Hamiltonian.
The value of $\Delta\rho({\cal E})$ at any complex energy ${{\cal E}=E-\frac{i}{2}\Gamma}$ is then given by the formula
\begin{equation}
\Delta\rho({\cal E})=\rho({\cal E})-\rho^{(0)}({\cal E}),
\label{Llev}
\end{equation}
where the two subtracted densities read as follows:
\begin{eqnarray}
\rho({\cal E})&=&\frac{1}{\pi}\sum_{k=1}^{N_{\vartheta}}\frac{-\frac{1}{2}(\Gamma\-\Gamma_k)+i(E-E_k)}{(E\-E_k)^2+\frac{1}{4}(\Gamma\-\Gamma_k)^2}
\nonumber\\
&+&\frac{1}{\pi}\sum_{l=N_{\vartheta}+1}^{\infty}\frac{-\frac{1}{2}(\Gamma\-\Gamma_l)+i(E-E_l)}{(E\-E_l)^2+\frac{1}{4}(\Gamma\-\Gamma_l)^2},
\label{Lrho}
\end{eqnarray}
\vspace{-3mm}
\begin{equation}
\rho^{(0)}({\cal E})=\frac{1}{\pi}\sum_{l=1}^{\infty}\frac{-\frac{1}{2}(\Gamma\-\Gamma^{(0)}_l)+i(E-E^{(0)}_l)}{(E\-E^{(0)}_l)^2+\frac{1}{4}(\Gamma\-\Gamma^{(0)}_l)^2}.
\label{Lrho0}
\end{equation}
The complex energies of discrete eigenstates of both Hamiltonians $\hat{{\cal H}}_{\vartheta}$ and $\hat{{\cal H}}^{(0)}_{\vartheta}$ are simple poles of the density \eqref{Llev}.
These are the energies ${\cal E}_k$ of identified resonances [the first line in Eq.\,\eqref{Lrho}], the energies ${\cal E}_l$ of $\hat{{\cal H}}_{\vartheta}$ on the rotated continuum ray [the second line in Eq.\,\eqref{Lrho}], and of the energies ${\cal E}^{(0)}_l$ of $\hat{{\cal H}}^{(0)}_{\vartheta}$ on the rotated continuum ray [Eq.\,\eqref{Lrho0}]. 
From the residue theorem we immediately see that a contour integral of $\Delta\rho({\cal E})$ along a closed loop in the complex plane ${\cal E}$ gives twice the difference between the number of eigenvalues of $\hat{{\cal H}}_{\vartheta}$ and $\hat{{\cal H}}^{(0)}_{\vartheta}$ inside the loop.

\begin{figure}[h!]
\includegraphics[width=\linewidth]{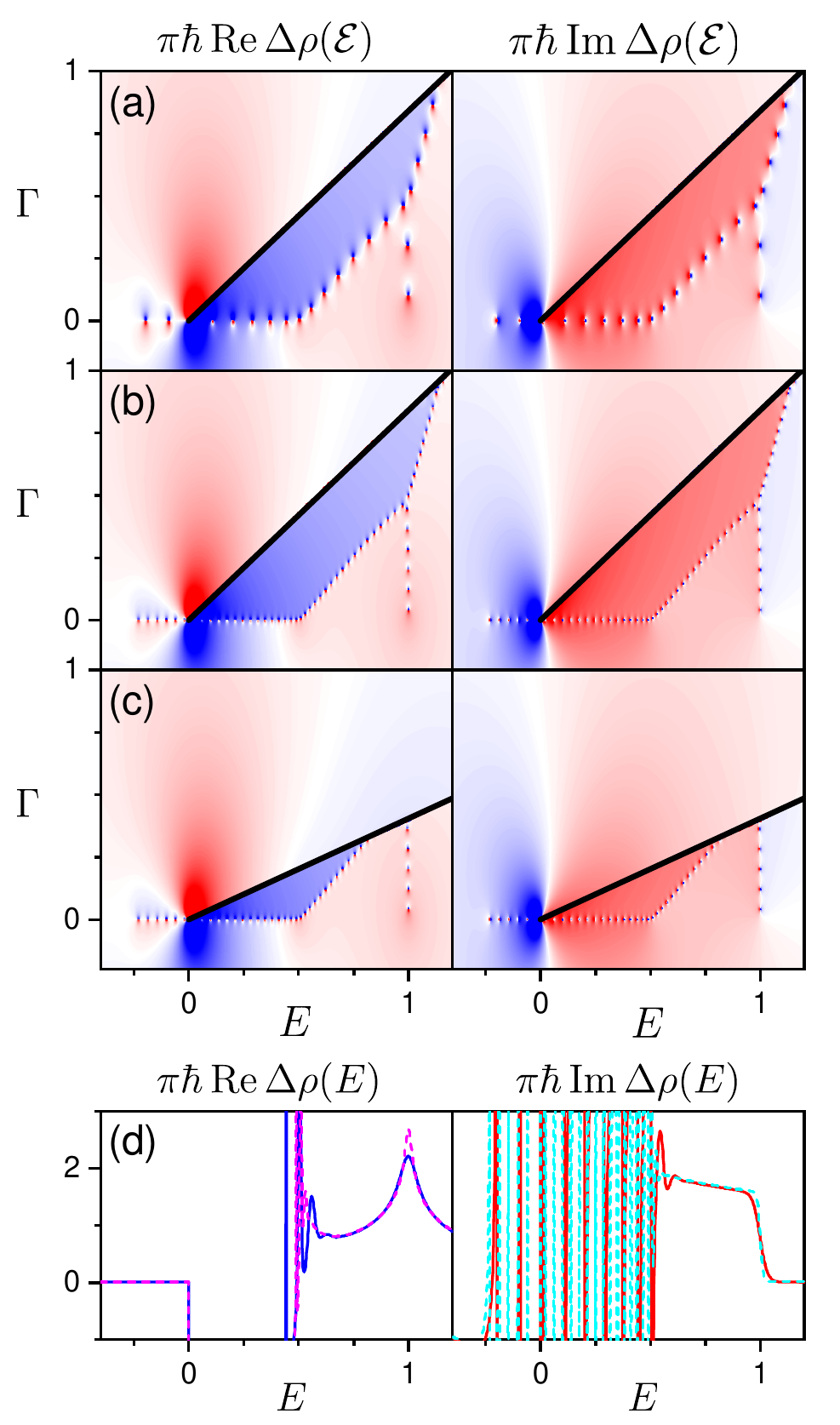}
\caption{
The complex continuum level density $\Delta\rho({\cal E})$ (real and imaginary parts in the left and right columns, respectively) for the potential from panel (f) of Fig.\,\ref{pots} calculated by the complex scaling method with different parameters.
Panels (a) and (b) compare densities for different size parameters~$\varkappa$, panels (b) and (c) those with different angles~$\vartheta$. 
Here $(\varkappa,\vartheta)$ is (a) $(20,0.2)$, (b) $(50,0.2)$, and (c) $(50,0.1)$. 
Panel (d) depicts densities with all the above parameter choices on the real energy axis.
The curves with different $\vartheta$ are indistinguishable, those with ${\varkappa=20}$ and 50 are drawn by the full and dashed lines, respectively. 
We observe independence of $\Delta\rho({\cal E})$ below the rotated continuum ray on the selected angle $\vartheta$ and an increase of the density of resonances and rotated-continuum states with $\varkappa$.
Note the real bound states with ${E<0}$ due to the deep potential minimum in Fig.\,\ref{pots}(f).
The densities in panel (d) can be compared to those in Fig.\,\ref{sin}(f).
All calculations are performed in a truncated basis of ${M=10^4}$ square-well eigenstates with ${L=150}$.
}
\label{varth}
\end{figure}

\begin{figure*}[t]
\includegraphics[width=\linewidth]{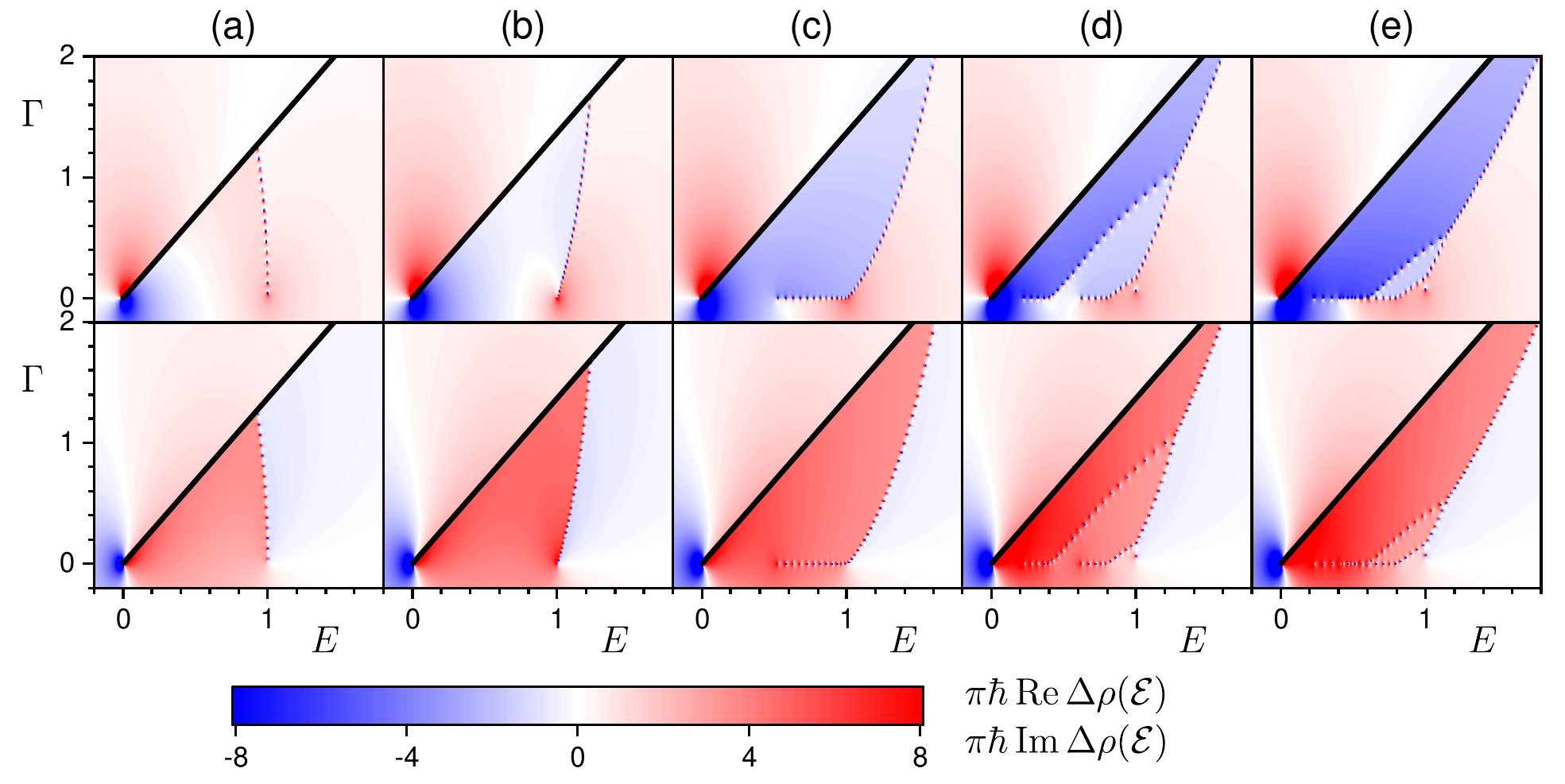}
\caption{
The complex continuum level density $\Delta\rho({\cal E})$ obtained by the complex scaling method for potentials from the respective panel (a)--(e) of Fig.\,\ref{pots}.
The real and imaginary parts of $\Delta\rho$ are in the upper and lower rows, respectively.
Note the chains of resonant states and their modifications at energies associated with the minima and maxima of the respective potentials.
The parameters are as follows: ${\varkappa=33.3}$, ${\vartheta=0.3}$, ${L=150}$, and ${M=10^4}$.
}
\label{dipoli}
\end{figure*}

The complex level density $\Delta\rho({\cal E})$ in the ${{\cal E}\in{\mathbb C}}$ plane for the six potentials from Fig.\,\ref{pots} is depicted in Figs.\,\ref{varth} and~\ref{dipoli}.
Figure~\ref{varth} shows $\Delta\rho({\cal E})$ calculated (with various choices of parameters $\vartheta$ and $\varkappa$) for the potential from panel (f) of Fig.\,\ref{pots}.
Note that this system contains also a discrete part of the spectrum, i.e., bound states with $E<0$.
Figure~\ref{dipoli} shows $\Delta\rho({\cal E})$ for the remaining five potentials from panels (a)--(e) of Fig.\,\ref{pots}.
The calculation is performed in the finite-$L$ approximation with a truncated basis including the lowest $M$ eigenstates of the infinite square-well Hamiltonian with ${x\in(-\frac{L}{2},+\frac{L}{2})}$.
All relevant parameters are specified in the captions.

Before focusing on the physical content of these figures, we comment on some technical aspects of the complex scaling method. 
These are illustrated in Fig.\,\ref{varth}.
It compares $\Delta\rho({\cal E})$ calculated with two choices of complex scaling angle $\vartheta$ and two values of the size parameter $\varkappa$.
The calculations in panels (b) and (c) for two angles $\vartheta$ from Eq.\,\eqref{Sim} verify the assumed independence of $\Delta\rho({\cal E})$ in the relevant part of the ${{\cal E}\in{\mathbb C}}$ plane on this parameter (see Sec.\,\ref{Coscame}).
The size of~$\vartheta$ is usually limited by computational constraints arising from the computer precision in the diagonalization of the truncated Hamiltonian matrix. 
However, as illustrated in Fig.\,\ref{varth}, its choice in well converged calculations (with a sufficiently large $L$) does not  influence the form of $\Delta\rho({\cal E})$ below the rotated continuum ray, particularly the positions of the disclosed resonances with ${\gamma_k<\vartheta}$.
The comparison of $\Delta\rho(E)$ on the real energy axis---see panel (d), where the curves for two different $\vartheta$ are indistinguishable---shows that the present finite-$L$ calculations are fully converged. 

In Fig.\,\ref{varth} we also compare the forms of $\Delta\rho({\cal E})$ obtained with two values of the size parameter $\varkappa$ from Eq.\,\eqref{size}.
In panels (a) and (b) we see that an increase of the system's size leads to an increase of the number of resonances and rotated-continuum states.
This is so because with increasing~$\varkappa$ both effective non-Hermitian Hamiltonians $\hat{{\cal H}}_{\vartheta}$ and $\hat{{\cal H}}^{(0)}_{\vartheta}$ head towards their classical limits with continuous spectra.
Since the higher-$\varkappa$ result is closer to the smooth classical limit, less additional smoothening is needed to extract the main energy dependence of $\Delta\rho(E)$ on the real energy axis.
We stress that the normalized density $\pi\hbar\,\Delta\rho(E)$ depicted in panel (d) does not show an overall increase with $\varkappa$, but only sharpening of the ESQPT structures discussed in Sec.\,\ref{Esqpts}.  
Let us note that the case of increasing~$\varkappa$, which affects all eigenstates of the effective Hamiltonians, must be distinguished from the case of increasing $L$, which affects only the rotated-continuum states. 

Now let us focus on some general features of the continuum level densities in Figs.\,\ref{varth} and \ref{dipoli}.
First we point out that the dependencies ${\rm Re}\,\Delta\rho({\cal E})$ and ${\rm Im}\,\Delta\rho({\cal E})$ in the ${{\cal E}\in{\mathbb C}}$ plane have a pictorial electrostatic interpretation.
It follows from the observation that the contribution of each discrete state to Eqs.\,\eqref{Lrho} and \eqref{Lrho0} has a form of a potential $v(\vecb{r})\propto(\vecb{d}\cdot\vecb{r})/r^2$ of the electric field generated by a dipole with moment $\vecb{d}$ in two-dimensional space, with $\vecb{r}$ standing for the coordinate vector that originates at the dipole position.
Indeed, considering the eigenenergy ${\cal E}_n$ (either a resonance, or a state belonging to the rotated continuum), we can associate the real and imaginary distance ${\rm Re}({\cal E}\-{\cal E}_n)$ and ${\rm Im}({\cal E}\-{\cal E}_n)$ of a selected energy ${\cal E}$ from ${\cal E}_n$ with the coordinate components $r_1$ and $r_2$ of the dipole analogy. 
The contribution of the $n$th state to ${\rm Re}\,\Delta\rho({\cal E})$ is proportional to ${-r_2/(\pi r^2)}$ and the contribution to ${\rm Im}\,\Delta\rho({\cal E})$ is proportional to ${r_1/(\pi r^2)}$.
This means that the real and imaginary parts of the continuum level density in the whole complex energy plane can be imagined as two distinct electric fields generated by a set of dipoles located at individual eigensolutions~${\cal E}_n$.
While the field ${\rm Re}\,\Delta\rho({\cal E})$ corresponds to dipole moments oriented antiparallel with the imaginary energy axis, the field ${\rm Im}\,\Delta\rho({\cal E})$ results from moments oriented parallel with the real energy axis.

Our main interest is focused on $\Delta\rho({\cal E})$ on the real energy axis ${\cal E}=E-i0$.
The real part of the level density is composed of sums of Cauchy peaks [see Eq.\,\eqref{led}] corresponding to individual eigensolutions of $\hat{{\cal H}}_{\vartheta}$ and $\hat{{\cal H}}^{(0)}_{\vartheta}$, 
\begin{eqnarray}
{\rm Re}\,\Delta\rho(E)=\sum_{k=1}^{N_{\vartheta}}C_{\Gamma_k}(E\-E_k)\ +
&& \!\!\!\sum_{l=N_{\vartheta}+1}^{\infty}\!\!\! C_{\Gamma_l}(E\-E_l)
\nonumber\\
-\sum_{l=1}^{\infty}&&C_{\Gamma^{(0)}_l}(E\-E^{(0)}_l).
\label{ccc}
\end{eqnarray}
Both ${\rm Re}\,\rho(E)$ and ${\rm Re}\,\rho^{(0)}(E)$ terms in this formula remind the standard density of the form $\sum_n\delta(E-E_n)$, where $n$ enumerates all eigenstates of the corresponding complex-scaled Hamiltonian, but each $\delta$-function is naturally smoothed with the aid of the actual width of the respective eigenstate.
The form of ${\rm Im}\,\Delta\rho(E)$ also follows Eq.\,\eqref{ccc}, but with the Cauchy peaks $C_{\Gamma}(E\-E_0)$ replaced by bipolar functions $B_{\Gamma}(E\-E_0)=2(E\-E_0)C_{\Gamma}(E\-E_0)/\Gamma$.

It can be anticipated that for real energies $E$ much larger than the maximal value of the potential $V(x)$, the eigensolutions ${\cal E}_l$ of the full Hamiltonian $\hat{{\cal H}}_{\vartheta}$ almost coincide with the eigensolutions ${\cal E}^{(0)}_l$ of the free Hamiltonian~$\hat{{\cal H}}^{(0)}_{\vartheta}$.
Therefore, their contributions to Eq.\,\eqref{Llev} approximately cancel out.
Moreover, even within the remaining low-energy terms, if $\vartheta$ is sufficiently large, the contributions to $\rho({\cal E})$ coming from the states at the rotated continuum ray [the second line of Eq.\,\eqref{Lrho}] have a tendency to approximately cancel out with the  contributions to $\rho^{(0)}({\cal E})$ on the rotated continuum ray [Eq.\,\eqref{Lrho0}].
The essential part of $\Delta\rho({\cal E})$ therefore comes from a finite number of resonant states located not too far from the real energy axis.
Wave functions of these states exhibit increased localization in the interaction region, their contributions thus survive the subtraction of the full and free Green operators in Eq.\,\eqref{lev}.
We stress that identification of the most important physical contributions is the main benefit of the complex scaling method.

However, if the angle $\vartheta$ is small, some of the relevant resonances may remain hidden.
Then the rotated continuum states play an important role and their contributions cannot mutually cancel.
The same conclusion holds also in a near vicinity of the point ${E=0}$, where the rotated continuum eigenstates of both the full and free Hamiltonians are close to the real energy axis and their contributions combine in a non-trivial way.
This is why the ${E\approx 0}$ region is the most difficult one with respect to the convergence of the finite-$L$ results to the correct limiting density $\Delta\rho(E)$.

\section{Complex extension of the tunneling time shift}
\label{Times}

\subsection{Eisenbud-Wigner time shift}
\label{EisWig}

Time relations in quantum scattering processes, and particularly time delays of the transmitted particle in quantum tunneling, got into the focus of theoretical interest already in early days of quantum mechanics \cite{Con31,Col32} and remain an important research topic up to the present days.
Theoretical analyses of this problem covering the time span of many decades can be found in Refs.\,\cite{Eis48,Wig55,Smi60,Tsa75,Lan94,Car02,Win06,Sok18} and the references therein.
The question which creates most fascination as well as controversy concerns the possibility of superluminal or even instantaneous occurrence of the transmitted particle on the exit from the tunneling potential.
At present, this question becomes a hot subject of experimental study by means of the attosecond metrology, see, e.g., Refs.\,\cite{Hen01,Sha12,Lan15,Sat19,Ram20}.

In the present work, we will employ the simplest definition of the tunneling time delay, the so-called Eisenbud-Wigner time \cite{Eis48,Wig55}.
It is determined from the energy variation of the phase shift $\varphi(E)$ of the transmission amplitude \eqref{tra}, namely by the formula
\begin{equation}
{\delta t}(E)=\hbar\frac{d}{dE}\varphi(E)=\pi\hbar\ \delta\rho(E),
\label{tis}
\end{equation}
where we used Eq.\,\eqref{dphi} to make an immediate link of ${\delta t}(E)$ to the real continuum level density $\delta\rho(E)$.
In case of a single resonance of centroid energy $E_k$ and width $\Gamma_k$, the time shift at ${E=E_k}$ is ${\delta t(E_k)=2\hbar/\Gamma_k}$, which is twice the average resonance lifetime, while far from the resonance center we have ${\delta t(E)\approx 0}$.

To understand the semiclassical meaning of the time delay \eqref{tis}, we need to avoid sharp resonant changes of the phase shift $\varphi(E)$ by using its smoothed form $\overline{\varphi}(E)$.
A~semiclassical estimate of the smoothed phase can be deduced from the Wentzel-Kramers-Brillouin (WKB) approximation, in which the transmitted wave at ${x=b}$ (end of the interaction region) is given by 
\begin{equation}
{\overline\beta}(E)\,e^{ipb/\hbar}= e^{ipa/\hbar}\ e^{i\left[\int\limits_{a}^{b}dx \sqrt{2m[E-V(x)]}+\phi\right]/\hbar},
\label{wkb}
\end{equation}
with $\phi$ denoting a constant that includes phase shifts at the classical turning points between allowed and forbidden regions \cite{Ber72}.
The formula \eqref{wkb} determines the smoothed phase $\overline{\varphi}(E)$ whose insertion into Eq.\,\eqref{tis} yields a smoothed time shift
\begin{equation}
\delta{\overline t}(E)=\!\!\!\!\underbrace{\int\limits_{\begin{smallmatrix}x\in[a,b]\\E\geq V(x)\end{smallmatrix}}\!\!\!\!\! 
dx\ \sqrt{\frac{m}{2[E\-V(x)]}}}_{{\overline t}^{(+)}(E)}-\underbrace{\sqrt{\frac{m}{2|E|}}\ (b-a)}_{\overline{t}^{(0)}(E)}.
\label{det}
\end{equation}
The integral in this expression is taken across all classically allowed (for given $E$) regions of the coordinate space between the points $a$ and $b$ demarcating the interaction domain (for multibarrier potentials these allowed regions can include several coordinate intervals).
This integral represents the time ${\overline t}^{(+)}(E)$ that a classical particle spends in the allowed regions during the passage from $a$ to $b$.
The subtracted term $\overline{t}^{(0)}(E)$ is the time of travel of a free particle across the full $(a,b)$ interval.
So we see that indeed, the Eisenbud-Wigner definition of the time shift includes no time delays resulting from the tunneling of the particle through the forbidden regions.

\subsection{Complex time shift}
\label{Comti}

Considering the full complex phase $\Phi(E)$ from Eq.\,\eqref{tra}, we introduce a complex time shift given by an analog of formula \eqref{tis}, namely
\begin{equation}
\Delta t(E)\equiv\hbar\frac{d}{dE}\Phi(E)=\pi\hbar\ \Delta\rho(E)
\label{cot}.
\end{equation}
From Eqs.\,\eqref{tauto} and \eqref{imdPhi} we immediately obtain 
\begin{eqnarray}
{\rm Re}\,{\Delta t}(E)&=&{\delta t}(E)
\label{tautoo},\\
{\rm Im}\,{\Delta t}(E)&=&-\frac{\hbar}{2}\frac{d}{dE}\ln|\beta(E)|^2
\label{itis}.
\end{eqnarray}
Using Eq.\,\eqref{wkb}, we write the smoothed forms of both real and imaginary time shifts in the pair of equations
\begin{eqnarray}
{\rm Re}\,\Delta{\overline t}(E)&=&\delta{\overline t}(E)={\overline t}^{(+)}(E)-\overline{t}^{(0)}(E),
\label{ret}\\
{\rm Im}\,\Delta{\overline t}(E)&=&\!\!\!\!\underbrace{\int\limits_{\begin{smallmatrix}x\in[a,b]\\E<V(x)\end{smallmatrix}}\!\!\!\!\!
dx\ \sqrt{\frac{m}{2[V(x)\-E]}}}_{\overline{t}^{(-)}(E)}
\label{imt},
\end{eqnarray}
where the first equation is just the formula \eqref{det}, with the integral involving all classically allowed spatial regions of the tunneling potential, while the second equation contains an integral across all classically forbidden regions.
The latter integral expresses the classical time $\overline{t}^{(-)}(E)$ which a particle with energy $-E$ would spent in these regions if the potential is inverted to $-V(x)$ so that the forbidden regions become allowed.
 
The expression in Eq.\,\eqref{imt} is related to complex-time solutions of the classical equations of motions inside the potential barriers, known from the instanton approach to the tunneling problem \cite{Par18}.
The standard instanton solution makes use of the Wick rotation $t\!\to\!-it$ of the time variable to derive the well-known WKB result on the tunneling probability \cite{Col77}.
With this transformation, the motion of the particle in the forbidden region becomes equivalent to the motion with energy $-E$ in an inverted potential $-V(x)$.
Related complex time approaches are used in the framework of the path integral description of tunneling processes, see, e.g., Refs.\,\cite{Lau72,Bal74,Col77,Mai97,Tak99,Tak01,Lev07,Deu10,Deu13,Tur14,Bra16,Har17}. 

Let us note that the extension of classical dynamics to the complex time represents an interesting, but not yet fully explored theoretical problem.
Once the time is considered complex, so must be the coordinate ${X\in{\mathbb C}}$ and momentum ${P\in{\mathbb C}}$ following from the complexified Hamilton equations. 
We assume that the complex time $T$ runs along a prescribed curve ${T(s)\in{\mathbb C}}$, where the parameterizing variable ${s\in{\mathbb R}}$ takes a role of a \uvo{proper time} of the moving particle.
The $s$-dependent trajectory in the complex phase space of $X$ and $P$ depends on the arbitrarily chosen curve $T(s)$.
In this sense, the complex-time dynamics is a generous broadening of the real-time dynamics.

\begin{figure}[t]
\includegraphics[width=\linewidth]{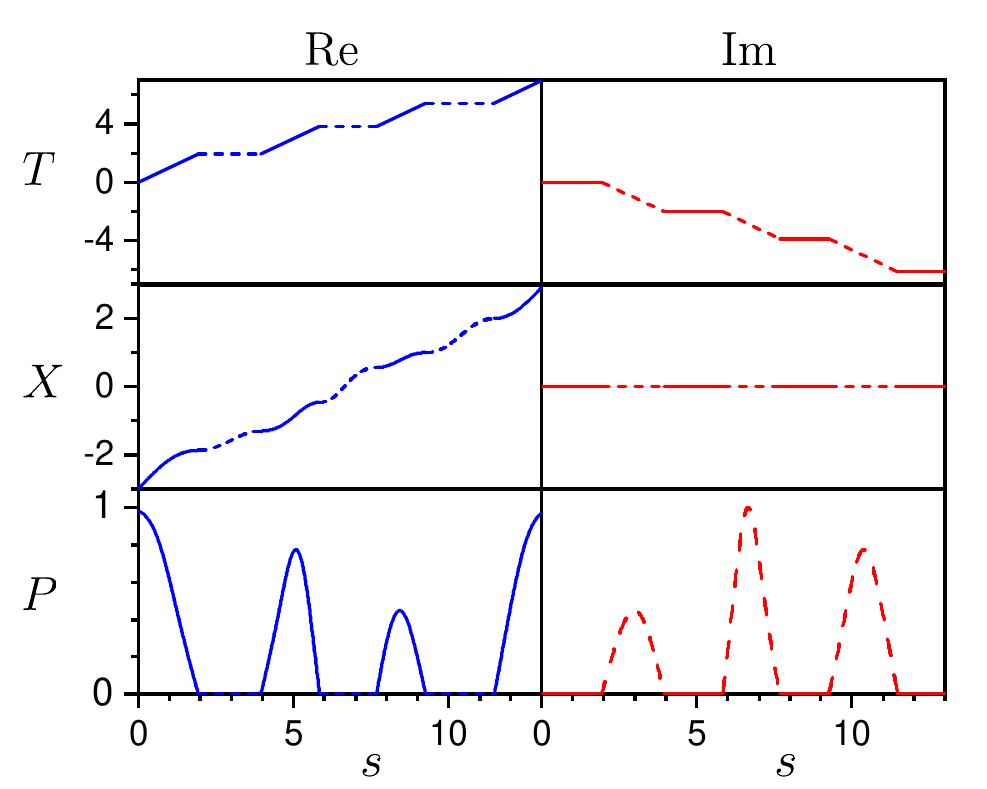}
\caption{
Example of a complex tunneling trajectory through the three-barrier potential in Fig.\,\ref{pots}(e) at energy ${E=0.5}$.
The segments corresponding to the allowed and forbidden regions are plotted by full and dashed lines, respectively. 
The dependence of complex time $T$, coordinate $X$ and momentum $P$ on the real running parameter $s$ are shown in the upper, middle and lower rows; the real and imaginary parts appear in the left and right columns.
Time runs along the real or imaginary axis in the allowed or forbidden regions, respectively, the coordinate remains real, and the momentum alternates between real and imaginary values.
This particular type of solutions of the complex Hamilton equations are derived in Appendix~\ref{AppB}. 
}
\label{traj}
\end{figure}

The results of this work highlight one specific class of very simple solutions of the complex Hamilton equations, which was discussed in connection with the multibarrier tunneling in Refs.\,\cite{Deu10,Deu13}.
The time $T(s)$ for an ${f=1}$ tunneling system is considered to run homogenously with variable $s$, but its direction in the complex plane  changes with respect to whether the particle is moving in the classically allowed or forbidden coordinate region.
In particular, ${\frac{d}{ds}T(s)=1}$ in the allowed region and ${\frac{d}{ds}T(s)=-i}$ in the forbidden region.
As a result, the complex time $T(s)$ with increasing $s$ accumulates in its real and imaginary part, respectively, the total traversal times of all allowed and all forbidden regions.
These times are calculated with the aid of potentials $V(x)$ and $-V(x)$. 
The evolution of coordinate remains real, ${X(s)=x(s)}$, but the momentum $P(s)$ switches between real and imaginary values in the allowed and forbidden regions, respectively.
An example of such a trajectory for the potential from Fig.\,\ref{pots}(e) is shown in Fig.~\ref{traj}, and the theoretical derivation is described in Appendix~\ref{AppB}.
If ${s\|{=}0}$ corresponds to the initial position ${x(0)\leq a}$ and momentum $P(0)\|{=}{\rm Re}\,P(0)\|{=}\sqrt{2mE}\|{>}0$ (i.e., the particle before entering the interaction region), the values of ${\rm Re}\,T(s)$ and ${\rm Im}\,T(s)$ for ${s>0}$ such that ${x(s)\geq b}$ (the particle after escaping the interaction region) correspond precisely to the values $\overline{t}^{(+)}(E)$ and $\overline{t}^{(-)}(E)$ in formulas \eqref{det} and \eqref{imt}.

\begin{figure}[t!]
\includegraphics[width=\linewidth]{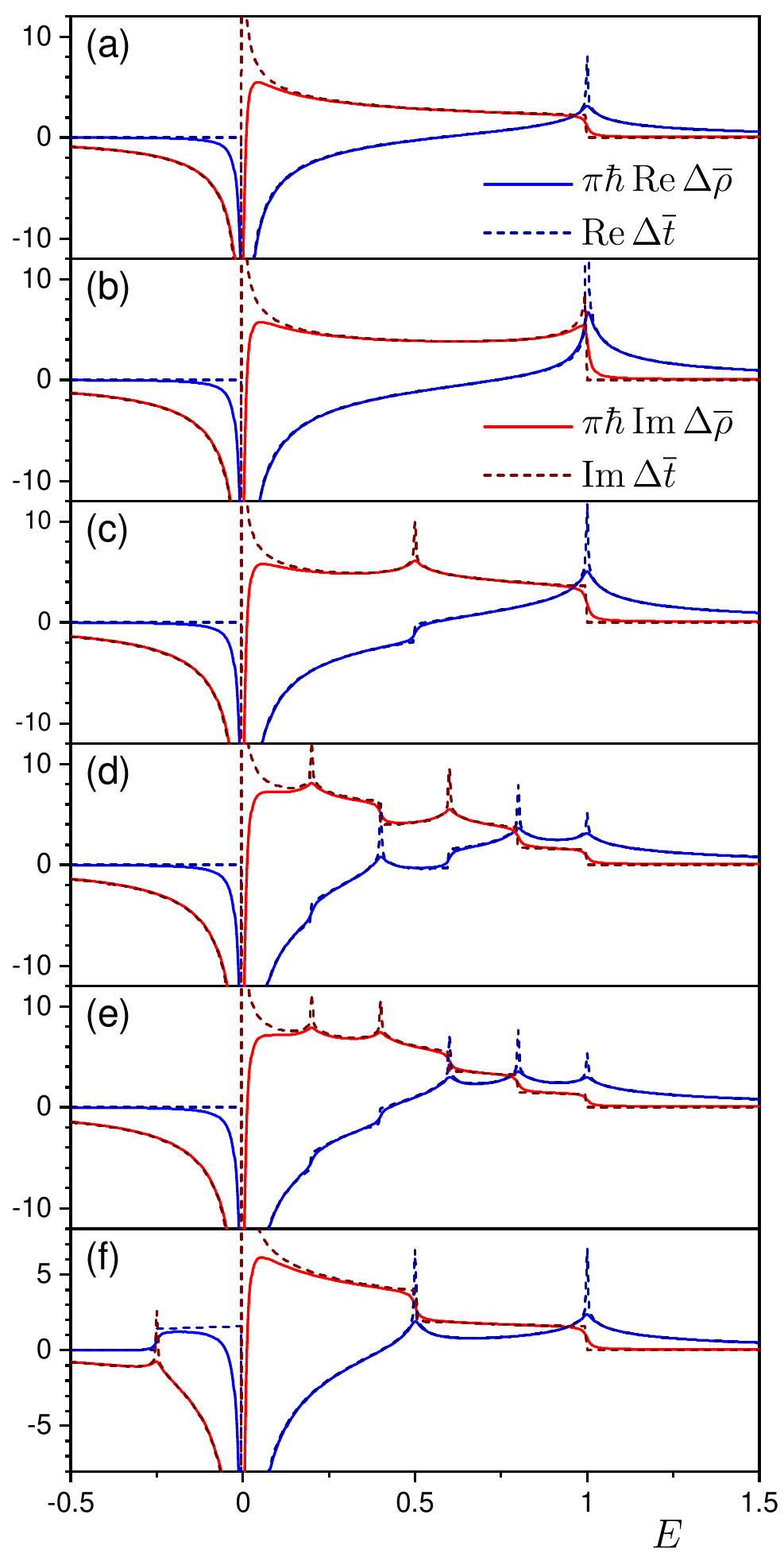}
\caption{The finite-size smoothed continuum level density ${\Delta\overline{\rho}(E)=\Delta\rho(E\|{+}i\epsilon)}$ and its infinite-size limit $\Delta\overline{t}(E)/{\pi\hbar}$ for the potentials from the respective panels (a)--(f) of Fig.\,\ref{pots}.
Parameters of the finite-size calculations are: ${\varkappa=200}$, ${\epsilon=0.01}$, ${\vartheta=0}$, ${L=150}$, ${M=10^4}$.
We observe finite-size precursors of the ESQPT-like singularities fully developed in the infinite-size dependencies. 
Note that in all panels we show a part of the ${E<0}$ domain, which is meaningful in panel (f) where we observe the presence of bound states.}
\label{sin}
\end{figure}

We stress that Eqs.\,\eqref{ret} and \eqref{imt} hold only for positive (physical) energies ${E>0}$.
Nevertheless, they can be formally extended also to ${E<0}$, where we can write:
\begin{eqnarray}
{\rm Re}\,\Delta{\overline t}(E)&=&{\overline t}^{(+)}(E),
\label{ret-}\\
{\rm Im}\,\Delta{\overline t}(E)&=&\overline{t}^{(-)}(E)-\overline{t}^{(0)}(E).
\label{imt-}
\end{eqnarray}
In this case, the \uvo{free propagation} belongs entirely to the forbidden region and its time $\overline{t}^{(0)}(E)$ contributes to ${\rm Im}\,\Delta{\overline t}(E)$.
So does a great part of the ${x\in(a,b)}$ motion in the potential, which generates the forbidden-region time  $\overline{t}^{(-)}(E)$.
The remaining allowed-region time $\overline{t}^{(+)}(E)$ contributing to ${\rm Re}\,\Delta{\overline t}(E)$ corresponds to the motion above possible negative minima of $V(x)$, cf.\,Fig.\,\ref{pots}(f).

Summarizing the above considerations, we write the final expression for the smoothed complex continuum level density in both ${E>0}$ and ${E<0}$ domains as follows: 
\begin{eqnarray}
\Delta{\overline\rho}(E)=\frac{\Delta{\overline t}(E)}{\pi\hbar}&=&
\underbrace{\frac{{\overline t}^{(+)}(E)\-\Theta(E)\,\overline{t}^{(0)}(E)}{\pi\hbar}}_{{\rm Re}\,\Delta{\overline\rho}(E)}
\nonumber\\
&+&i\ \underbrace{\frac{{\overline t}^{(-)}(E)\-\Theta(-E)\,\overline{t}^{(0)}(E)}{\pi\hbar}}_{{\rm Im}\,\Delta{\overline\rho}(E)}.\quad
\label{deti}
\end{eqnarray}
Here the times ${\overline t}^{(+)}(E)$, ${\overline t}^{(-)}(E)$ and ${\overline t}^{(0)}(E)$ are calculated from the expressions in Eqs.\,\eqref{det} and \eqref{imt}, and $\Theta$ is a step function (\,=\,0 or 1 for negative or semipositive arguments, respectively).
The formula \eqref{deti} provides a simple semiclassical estimate of both the real and imaginary parts of the smoothed continuum level density.
We stress its apparent similarity to the relation \eqref{Tper} between the smoothed level density of an ${f=1}$ bound system and the period $\tau(E)$ of classical orbits at energy $E$.
The denominators in these formulas differ by factor 2 because the time shifts in Eq.\,\eqref{deti} include only a half of the return trajectory.
A new aspect of the present situation is the duality following from the existence of the allowed and forbidden regions of the classical motions, which contribute to the real and imaginary parts of the continuum level density.
This will turn important in the description of singularities of $\Delta{\overline\rho}(E)$ connected with stationary points of $V(x)$, as discussed in Sec.\,\ref{Esqpts}.

A comparison of the smoothed complex continuum level density calculated for finite-size tunneling systems with the semiclassical estimate based on the formula \eqref{deti} is presented in Fig.\,\ref{sin}.
Panels (a)--(f) depict results for the six tunneling potentials from the respective panels of Fig.\,\ref{pots}.
The semiclassical estimate corresponds to ${\varkappa\to\infty}$, while the quantum calculation was performed with the size parameter ${\varkappa=200}$.
Comparing in each panel of Fig.\,\ref{sin} the curve $\pi\hbar\,{\rm Re}\,\Delta{\overline\rho}(E)$ with ${\rm Re}\,\Delta{\overline t}(E)$, and the curve $\pi\hbar\,{\rm Im}\,\Delta{\overline\rho}(E)$ with ${\rm Im}\,\Delta{\overline t}(E)$, we confirm a satisfactory overall agreement of the observed finite-size behavior with the corresponding infinite-size limit.
The curves are not compatible only in a vicinity of ${E=0}$.
Even in this most difficult region (see end of Sec.\,\ref{Coscale}) our results are well converged, however the discrepancies between the finite-size curves and the semiclassical curves (which at ${E=0}$ show divergencies of both signs) appear as an artifact of the smoothening procedure.

\section{ESQPT-like tunneling singularities}
\label{Esqpts}

\begin{table*}[t]
\begin{ruledtabular}
\begin{tabular}{lcc}
type of stationary point of $V(x)$ & ${\rm Re}\,\Delta{\overline\rho}_{\rm irr}(E)\propto$ & ${\rm Im}\,\Delta{\overline\rho}_{\rm irr}(E)\propto$ 
\vspace{1mm}\\ \hline \vspace{-2mm}\\
$n=2$ minimum & $\Theta(E-E_0)$ & $\ln|E_0-E|^{-1}$ \\
$n=2$ maximum & $\ln|E-E_0|^{-1}$ & $\Theta(E_0-E)$ \\
$n=4,6,...$ minimum & $|E-E_0|^{-(n-2)/2n}\Theta(E-E_0)$ & $|E_0-E|^{-(n-2)/2n}$ \\
$n=4,6,...$ maximum & $|E-E_0|^{-(n-2)/2n}$ & $|E_0-E|^{-(n-2)/2n}\Theta(E_0-E)$ \\
$n=3,5,7...$ saddle point & $|E-E_0|^{-(n-2)/2n}$ & $|E_0-E|^{-(n-2)/2n}$ \\
square well or barrier & $|E-E_0|^{-1/2}\Theta(E-E_0)$ & $|E_0-E|^{-1/2}\Theta(E_0-E)$ \\
\end{tabular}
\end{ruledtabular}
\caption{Irregular components (up to multiplicative factors) of the real and imaginary parts of the ${\varkappa\to\infty}$ continuum level density near energy $E_0$ corresponding to various types of stationary points of the tunneling potential from Eq.\,\eqref{stapot}.}
\label{tab}
\end{table*}

Since the relation \eqref{deti} between the continuum level density and time shift in unbound tunneling systems fully parallels relation~\eqref{Tper} between the level density and period in bound systems, it must encode singularities of continuous spectra similar to the ESQPT singularities of discrete spectra.
These continuum analogues of ESQPTs are most commonly caused by stationary points of the potential $V(x)$, which generate divergencies or other non-analyticities of the real and imaginary time shifts. 
Here we introduce a general typology of these non-analyticities for stationary points that allow a local Taylor expansion of $V(x)$.
Effects resulting from non-analytic minima of maxima of the potential (such as the $\vee$ or $\wedge$ shaped ones) can be derived as well, but we do not discuss them here. 

Let $V^{(\pm)}(x)$ denotes the normal and inverted potentials $\pm V(x)$, where the sign $+$ or $-$ applies in the allowed or forbidden regions, respectively, and $x_0$ is a stationary point of $V^{(\pm)}(x)$ such that the lowest power of the expansion in ${\delta x=x-x_0}$ is an integer ${n\geq 2}$. 
So 
\begin{equation}
V^{(\pm)}(x)\stackrel{\rm loc}{=}
E_0+c\,(\delta x)^n+O\left((\delta x)^{n+1}\right)
\label{stapot},
\end{equation}
where $E_0$ is the stationary point energy and $c$ is an arbitrary constant.
The note \uvo{loc} above the equality sign emphasizes its only local validity near $x_0$.
If ${n=2,4,6,...}$, the stationary point is a minimum for ${c>0}$ or a maximum for ${c<0}$, and if ${n=3,5,7,...}$, the stationary point is a saddle.

The times ${\overline t}^{(\pm)}(E)$ for energies close to the stationary point energy $E_0$ can be split into a regular component ${\overline t}^{(\pm)}_{\rm reg}(E)$, which depends on the dynamics away from the stationary point and is a smooth function of energy, and an irregular component ${\overline t}^{(\pm)}_{\rm irr}(E)$, which depends on the motion close to the stationary point and has a non-analyticity at ${E=E_0}$.
We are now interested only in the irregular component as it determines the type of singularity of the continuum level density. 
It is given by the time accumulated in a certain coordinate interval ${x\in(x_0\-\ell,x_0\+\ell)}$ around the stationary point (with ${\ell>0}$ denoting an arbitrary small distance) and can be determined from the integral
\begin{eqnarray}
{\overline t}^{(\pm)}_{\rm irr}(E)=
\sqrt{\frac{m}{2}}\int\limits_{x_0-\ell}^{x_0+\ell}
dx\ \frac{\Theta(E\-V^{(\pm)}(x))}{\sqrt{E\-V^{(\pm)}(x)}}
\nonumber\\
\propto
|\delta E|^{-(n-2)/2n}
\!\!\!\!\!\!
\int\limits_{0}^{\ell(|c|/|\delta E|)^{1/n}}
\!\!\!\!\!\!
dq\ \frac{\Theta(1\-\sigma q^n)}{\sqrt{1\-\sigma q^n}}.
\label{int}
\end{eqnarray}
Here, ${\delta E=E\-E_0}$ and ${\sigma={\rm sgn}(c\,\delta E)}$.
The energy factor in front of the integral in the second line of Eq.\,\eqref{int} captures the leading-order diverging term of ${\overline t}^{(\pm)}_{\rm irr}(E)$ at the stationary point energy if the integral has a regular Taylor expansion in $(\delta E)^{1/2n}$.
This condition is not satisfied for ${n=2}$ with ${c<0}$, but in this case the problem can be solved by other means.
A more detailed analysis of singularities caused by stationary points can be found in Refs.\,\cite{Str14,Str16}.

The effects of various stationary points of the normal or inverted tunneling potential $V^{(\pm)}(x)$ on the semiclassical time shift 
can be summarized as follows:
A local maximum of any even power $n$ gives rise to two possible types of singularity,
\begin{equation}
{\overline t}^{(\pm)}_{\rm irr}(E)
\propto\left\{\!\!
\begin{array}{ll}
\ln|E-E_0|^{-1}&{\rm for\ }n\=2,\\
|E-E_0|^{-(n-2)/2n}&{\rm for\ }n\=4,6,...
\end{array}\right.
\label{mini}
\end{equation}
We note that the proportionality constant can be different on the both ${\delta E<0}$ and  ${\delta E>0}$ sides of the singularity.
In the limiting case of ${n\to\infty}$, which characterizes the square potential barrier from Fig.\,\ref{recta}, the contribution in the second line of Eq.\,\eqref{mini} is nonzero only for ${\delta E\geq 0}$, so we obtain ${{\overline t}^{(\pm)}_{\rm irr}(E)\propto\Theta(E-E_0)/\sqrt{E\-E_0}}$.
Similarly, a local minimum of any power ${n=2,4,6,...}$ leads to
\begin{equation}
{\overline t}^{(\pm)}_{\rm irr}(E)
\propto|E-E_0|^{-(n-2)/2n}\,\Theta(E\-E_0)
\label{maxi}.
\end{equation}
Note that for a quadratic potential minimum with ${n=2}$ this formula predicts just an upward jump of the time shift at ${E=E_0}$, while for the square well with ${n\to\infty}$ it leads to the square-root divergence of the time shift on the ${E\geq E_0}$ side, as in the case of the square barrier.
Finally, a saddle point of power ${n=3,5,7...}$ yields
\begin{equation}
{\overline t}^{(\pm)}_{\rm irr}(E)
\propto|E-E_0|^{-(n-2)/2n}
\label{sadd}
\end{equation}

As follows from Eq.\,\eqref{deti}, any singularity in the tunneling time generates the same kind of singularity in the smoothed continuum level density $\Delta{\overline\rho}(E)$.
All possible types of singularities resulting from stationary points of the form \eqref{stapot} are summarized in Table~\ref{tab}.
We emphasize the duality of singularities in the real and imaginary parts of $\Delta{\overline\rho}(E)$ due to the ${E\to-E}$ and ${V(x)\to -V(x)}$ inversion connected with the instanton-like semiclassical solutions.
So, for instance, a quadratic maximum in the tunneling potential $V(x)$ generates a logarithmic divergence of the real density ${\rm Re}\,\Delta{\overline\rho}(E)$ and a downward step discontinuity of the imaginary density ${\rm Im}\,\Delta{\overline\rho}(E)$.
These effects are reversed in case of a quadratic minimum, which generates an upward step discontinuity of the real density and a logarithmic divergence of the imaginary density.

The singularities listed in Table~\ref{tab} represent continuum analogues of the ESQPT singularities in ${f=1}$ bound systems.
The original ESQPTs are caused by stationary points of bound classical dynamics and affect the smoothed (semiclassical) level density describing discrete energy spectra \cite{Cej06,Cap08,Bra13,Lar13,Str14,Bas14a,Bas14,Rel14,Kop15,Pue16,Str16,Sin17,Byc18,Gar18,Kha19,Mac19,Cej20}.
They emerge as non-analyticities of a normalized exact level density in the system's infinite-size limit. 
The present continuum form of the ESQPT affects both real and imaginary parts of the smoothed continuum level density, i.e., the exact real and imaginary densities in the ${\varkappa\to\infty}$ limit.
It therefore represents a direct analogue and a dual extension of the original ESQPT concept.

The ESQPT-like singularities of the complex continuum level density for the potentials from Fig.\,\ref{pots} can be seen in Fig.\,\ref{sin}.
The true singularities (jumps and logarithmic or power-law divergences) are present only in the semiclassical (${\varkappa\to\infty}$) curves $\Delta{\overline t}(E)$, while the quantum (finite-$\varkappa$) calculations of $\Delta{\overline\rho}(E)$ demonstrate only some precursors of the predicted non-analytic behavior.
In accord with the above explanations, the stationary points of $V(x)$ produces different types of singularities in the components ${\rm Re}\,\Delta{\overline\rho}(E)$ and ${\rm Im}\,\Delta{\overline\rho}(E)$.
In panels (a) and (b), respectively, we find singularities associated with a quadratic (${n=2}$) and quartic (${n=4}$) maxima (or minima) of $V^{(+)}(x)$ [or $V^{(-)}(x)$] at energy ${E_0=1}$.
Panels (c)--(f) contain more singularities associated with quadratic maxima and minima of the respective potentials $V^{(\pm)}(x)$.
In particular, panels (c) and (f) show two singularities at ${E_0=0.5}$ and 1, while panels (d) and (e) have five singularities at energies ${E_0=0.2}$, 0.4, 0.6, 0.8 and 1.
The observed behavior is in all cases fully consistent with the results listed Table~\ref{tab}.

\begin{figure}[t]
\includegraphics[width=\linewidth]{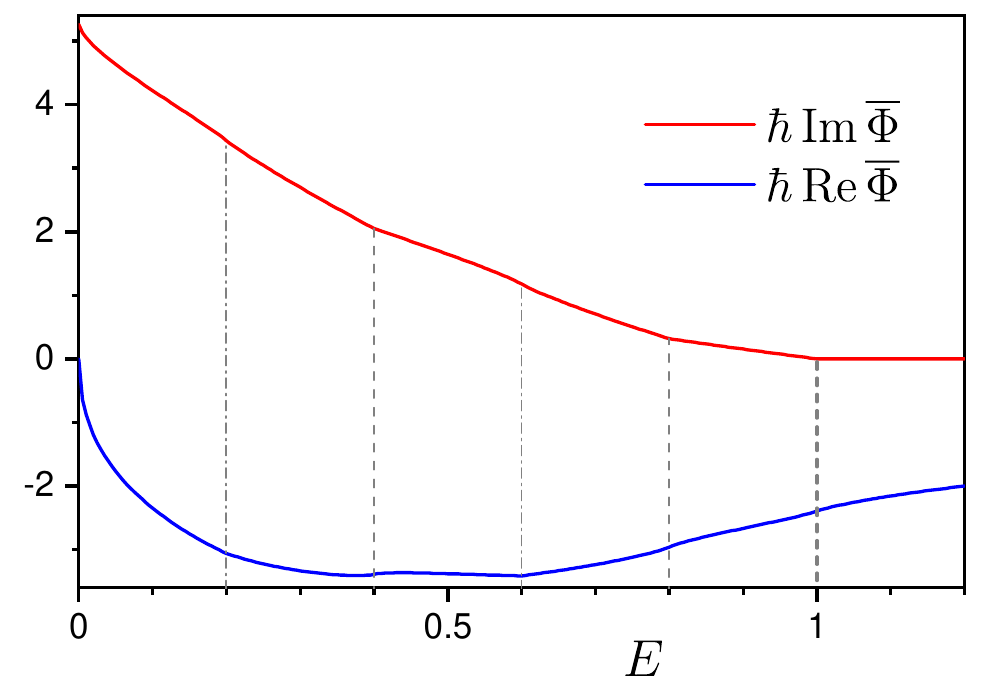}
\caption{
Real and imaginary parts of the smoothed phase ${\overline\Phi}(E)$ obtained via Eqs.\,\eqref{dPhi} and \eqref{deti} for a ${\varkappa\to\infty}$ system with the potential from Fig.\,\ref{pots}(d).
The real part is determined up to an additional constant which is set so that ${{\rm Re}\,{\overline\Phi}(0)=0}$. 
Energies of the stationary points (marked by the vertical lines) locate non-analyticities of ${\overline\Phi}(E)$. 
}
\label{phase}
\end{figure}

We stress that the continuum ESQPT-like singularities lead to observable consequences in the form of the transmission amplitude.
In particular, as follows from Eq.\,\eqref{dPhi}, the singularities from Table~\ref{tab} appear in the infinite-size limit of the first derivative $\frac{d}{dE}\Phi(E)$ of the tunneling phase.
An example of the ${\varkappa\to\infty}$ ESQPT-induced structures in ${\rm Re}\,\overline{\Phi}(E)$ and ${\rm Im}\,\overline{\Phi}(E)$ for the potential from Fig.\,\ref{pots}(d) is presented in Fig.\,\ref{phase}.
We know that the level densities ${\rm Re}\,\Delta{\overline\rho}(E)$ and ${\rm Im}\,\Delta{\overline\rho}(E)$ for the given potential manifest logarithmic divergences and jumps at energies ${E_0=0.2}$, 0.4, 0.6, 0.8 and 1, see Fig.\,\ref{sin}(d), so at the same energies the respective phases ${\rm Re}\,\overline{\Phi}(E)$ and ${\rm Im}\,\overline{\Phi}(E)$ show little step-like structures (with locally vertical tangents) and breaks.

Let us note that the anomalies connected with local maxima of the tunneling potential were identified and semiclassically analyzed in the context of chemical physics already in Refs.\,\cite{Con68+,Roy78}. 
Here we extend these results to general stationary points of arbitrary tunneling potentials and connect them to the more elaborated description of ESQPTs in bound systems.

\section{Conclusions}
\label{Clouseau}

This follow-up paper of our previous publication~\cite{Str20} provides a more detailed account of the complex-extended continuum level density in one-dimensional tunneling problems and its semiclassical interpretation.
The larger space allows us to present a more complete and transparent analysis, including explicit derivation of the key surmises of Ref.\, \cite{Str20}.
We verify our theoretical conclusions by more numerical examples, employing additional sample potentials.

The main result of our analysis is the evidence that the semiclassical formulation of the tunneling problem with complex-extended time yields correct estimates of the smoothed continuum level density and the smoothed transmission amplitude.
These estimates become exact in the ${\varkappa\to\infty}$ limit.
Let us stress that according to a common opinion, any semiclassical formulation of quantum tunneling is regarded as an oxymoron.
Here we deepen the understanding (proposed and elaborated in the previous literature on the instanton solutions and generalized path integrals)
that both opposites can be married through the concept of complex time.
In particular, the simple type of complex-time tunneling trajectories exemplified in Fig.\,\ref{traj} and analyzed in Appendix~\ref{AppB} turn out to be the key for describing the observed behavior of the smoothed transmission amplitudes.

An immediate consequence of our analysis is the generalization of the ESQPT singularities to systems with continuous energy spectra associated with scattering phenomena.
As in bound systems with discrete spectra, the singularities of the continuum level density follow from the existence of classical stationary points of the potential $V(x)$.
Dual ESQPT structures in the real and imaginary parts of the continuum level density, which are based on potentials $V(x)$ and $-V(x)$ and on the existence of classically allowed and forbidden spatial regions, represent an interesting enrichment of the ESQPT phenomenology.
Let us note that the generalized ESQPT structures in the continuum level density associated with local minima, maxima or saddles of the potential indicate abrupt qualitative changes of the tunneling trajectories for the particle energy crossing the critical value ${E_0=V(x_0)}$.
This results in non-analytic evolutions of the tunneling observables in the ${\varkappa\to\infty}$ limit.
In this sense the ESQPT structures indeed constitute a specific kind of dynamical critical effect.

From the theoretical point of view, an apparent open question concerns the extension of the present ${f=1}$ results to systems with ${f>1}$.
The ESQPTs in closed systems with more than one degrees of freedom appear as discontinuities or divergencies in a higher, typically the ${(f\-1)}$th derivative of the level density \cite{Str16}, and the same is anticipated for the continuum systems.
Such generalization can be elaborated explicitly for separable systems, either of the spherically symmetric or Cartesian type. i.e. systems with potentials ${V(\vecb{x})=V(|\vecb{x}|)}$ or ${V(\vecb{x})=\sum_{i=1}^{f}V_i(x_i)}$, respectively.
This might be a starting point of a more general future analysis including non-separable continuum systems.

The ESQPT non-analyticieties in continuum systems can be experimentally detected in feasible tunneling experiments.
Verification of such effects in both the real and imaginary phase of the transmission amplitude will require to use an interference setup, in which both the intensity and phase of the wave transmitted through a given potential can be compared with those of a freely propagated wave.
We conclude by noting that the present day nanotechnology makes it possible to synthesize diverse resonant tunneling potentials (see, e.g., Refs.\,\cite{See01,Bha06,Suz10,Bri13,Gol15,Tao19}), so that experimental tests and perhaps even some applications of the above-explained concepts may be a matter of near future.


\section*{Acknowledgments}

We thank Michal Kloc for useful comments on the manuscript.
The work was supported by the Czech Science Foundation (grant nos.\,20-09998S and 20-21179S) and by the Charles University (project UNCE/SCI/013).

\appendix

\section{Continuum level density and the transmission amplitude} 
\label{AppA}

Here we derive the relation \eqref{dPhi} between the complex continuum level density and the complex phase of the transmission amplitude.
The level density is defined by Eq.\,\eqref{lev}, where the trace can be performed in any complete basis set. 
We use the $\delta$-normalized position eigenstates $\{|x\ra\}_{x \in {\mathbb R}}$, so $\Delta \rho({\cal E})$ on the real energy axis reads
\be \label{Delta-rho-x-basis}
   \Delta \rho(E)=
   \frac{i}{\pi}\int_{-\infty}\limits^{+\infty}\!\!\left( G(E;x,x)\|{-}G^{(0)}(E;x,x) \right){\rm d}x.
\ee
The Green functions $G(E;x,x')\|{=}\matr{x}{\hat{G}(E\|{+}i0_+)}{x'}$ and $G^{(0)}(E;x,x')\|{=}\matr{x}{\hat{G}^{(0)}(E\|{+}i0_+))}{x'}$ (here the retarded Green functions) are analyzed in standard textbooks \cite{Tay00,Fri13}.
Importantly, $G(E;x,x')$ is expressible in terms of the two degenerate eigenstates $\psi_{E\pm}(x)$ of $\hat{H}$ corresponding to the same continuum energy level ${E=p^2/(2m)>0}$:
\be \label{G-E-+-x-x}
   G(E;x,x)=-i\,\frac{m}{\hbar p}\ \frac{\psi_{E-}(x)\ \psi_{E+}(x)}{\beta(E)}.
\ee
These states 
exhibit the following asymptotic behavior:
\begin{eqnarray}
         \label{tilde-psi-E-+-bc-b}
         \psi_{E+}(x\to+\infty) &=& \beta_+(E)e^{+ipx/\hbar},\\
         \label{tilde-psi-E-+-bc-a}
         \psi_{E+}(x\to-\infty) & = & e^{+ipx/\hbar}+\alpha_+(E)e^{-ipx/\hbar},\\
         \label{tilde-psi-E---bc-b}
         \psi_{E-}(x\to+\infty) & = & e^{-ipx/\hbar}+\alpha_-(E)e^{+ipx/\hbar},\\
         \label{tilde-psi-E---bc-a}
         \psi_{E-}(x\to-\infty) & = & \beta_-(E)e^{-ipx/\hbar}.
\end{eqnarray}
While the boundary conditions (\ref{tilde-psi-E-+-bc-b}) and (\ref{tilde-psi-E-+-bc-a}) coincide with Eq.\,\eqref{asy} and describe the particle of energy $E$ approaching the interaction region from the left-hand side, the conditions (\ref{tilde-psi-E---bc-b}) and (\ref{tilde-psi-E---bc-a}) describe an analogous setup with the particle approaching from the right-hand side.  
The transmission amplitudes satisfy the relations $\beta_+(E)\|{=}\beta_-(E)\|{\equiv}\beta(E)$ following from the time-reversal symmetry, but the reflection amplitudes $\alpha_+(E)$ and $\alpha_-(E)$ may differ by a phase factor.
For the Green function of the free Hamiltonian $\hat{H}^{(0)}$ we get
\be \label{G-0-E-+-x-x}
         G^{(0)}(E;x,x)=-i \frac{m}{\hbar p}.
\ee      
A straightforward combination of (\ref{Delta-rho-x-basis}), (\ref{G-E-+-x-x}) and (\ref{G-0-E-+-x-x}) yields an intermediate result
\be \label{Delta-rho-x-basis-intermediate}
	\Delta \rho(E)=\frac{m}{\pi\hbar p}\int\limits_{-\infty}^{+\infty}\left[\frac{\psi_{E-}(x)\psi_{E+}(x)}{\beta(E)}-1\right]dx,
\ee
which will be further simplified by using the boundary conditions (\ref{tilde-psi-E-+-bc-b})--(\ref{tilde-psi-E---bc-a}).

The first step consists in evaluating an integral
\be
	\label{inta}
        I_L(E,E')=\int\limits_{-L}^{+L}\psi_{E'-}(x)\psi_{E+}(x)dx,
\ee
where ${L>0}$ is a fixed parameter and ${E,E'>0}$ are two energies corresponding to momenta $p,p'$.
We have
\begin{eqnarray} 
\label{TISCHE-E-(+1)}
-\frac{\hbar^2}{2m}\frac{\partial^2\psi_{E+}(x)}{\partial x^2}+V(x)\psi_{E+}(x)&=&E\psi_{E+}(x),\qquad\\
\label{TISCHE-E'-(-1)}
-\frac{\hbar^2}{2m}\frac{\partial^2\psi_{E'-}(x)}{\partial x^2}+V(x)\psi_{E'-}(x)&=&E'\psi_{E'-}(x).
\end{eqnarray}
If we multiply Eq.\,(\ref{TISCHE-E-(+1)}) by $\psi_{E'-}(x)$, subtract Eq.\,(\ref{TISCHE-E'-(-1)}) multiplied by $\psi_{E+}(x)$, and apply $\int_{-L}^{+L}dx$, we obtain
\begin{eqnarray} 
	\label{I-L-take-1}
	&& \frac{2m}{\hbar^2}(E-E')I_L(E,E')=\\
	&&
	\int\limits_{-L}^{+L}\left[\psi_{E+}(x)\frac{\partial^2\psi_{E'-}(x)}{\partial x^2}-\psi_{E'-}(x)\frac{\partial^2\psi_{E+}(x)}{\partial x^2}\right]dx.
	\nonumber
\end{eqnarray}
Turning the energies into momenta and integrating in the second line {\it per partes}, we get
\begin{eqnarray} 
	\label{I-L-take-2}
	&& \frac{(p+p')(p-p')}{\hbar^2}I_L(E,E')= \\
	&& \left[\psi_{E+}(x)\frac{\partial\psi_{E'-}(x)}{\partial x}-\psi_{E'-}(x)\frac{\partial\psi_{E+}(x)}{\partial x}\right]_{x=-L}^{x=+L}. 
	\nonumber
\end{eqnarray}
For large $L$, the right-hand side of this expression can be evaluated by inserting Eqs.\,(\ref{tilde-psi-E-+-bc-b})--(\ref{tilde-psi-E---bc-a}):
\begin{eqnarray} 
	\cdots=i\frac{p+p'}{\hbar}\left(\beta(E')e^{-i(p-p')L/\hbar}-\beta(E)e^{i(p-p')L/\hbar}\right)
	\nonumber\\
	-i\frac{p-p'}{\hbar}\bigl(\alpha_+(E)\beta(E')+\alpha_-(E')\beta(E)\bigr) e^{i(p+p')L/\hbar}\qquad
	\label{I-L-take-3}
\end{eqnarray}
Approaching the limits ${E'\to E}$ and ${L\to\infty}$, we see that the expression in the second line oscillates very quickly with varying energy and becomes insignificant after averaging over an arbitrarily small energy interval. 
Hence near these limits we can write 
\be
\frac{(p\|{-}p')}{\hbar}I_L(E,E')\|{=}i\beta(E')e^{-i(p-p')L/\hbar}\|{-}\beta(E)e^{i(p-p')L/\hbar}.
\ee
Finally, expressing $\beta(E')$ through its Taylor expansion,
\be
	\beta(E')=\beta(E)+\frac{d\beta(E)}{dE}(E'\|{-}E)+O\bigl((E'\|{-}E)^2\bigr),
\ee
we get an outcome for the integral \eqref{inta},
\begin{eqnarray}
I_L(E,E')&=&2\beta(E)\,\frac{\hbar\,\sin{\frac{(p-p')L}{\hbar}}}{p-p'}
\label{I-L-take-5}
\\
&-&i\,\frac{d\beta(E)}{dE}\,\frac{\hbar(p\|{+}p')}{2m}\,e^{-i(p-p')L/\hbar}+O(p\|{-}p'),
\nonumber
\end{eqnarray}
which for ${E=E'}$ yields:
\be
\label{I-L-take-6}
I_L(E,E)=2L\beta(E)-i\,\frac{\hbar p}{m}\,\frac{d\beta(E)}{dE}.
\ee

In the second quick step, we substitute the last formula (\ref{I-L-take-6}) to Eq.\,(\ref{Delta-rho-x-basis-intermediate}) and obtain
\begin{eqnarray}
&&\int\limits_{-\infty}^{+\infty}\left[\frac{\psi_{E-}(x)\psi_{E+}(x)}{\beta(E)}-1\right]dx
=\lim\limits_{L\to\infty}\left(\frac{I_L(E,E)}{\beta(E)}\|{-}2L\right)
\nonumber\\
&&=-i\,\frac{\hbar p}{m}\,\frac{\frac{d\beta(E)}{dE}}{\beta(E)}
\quad\Rightarrow\quad
\Delta \rho(E)=-\frac{i}{\pi}\,\frac{\frac{d\beta(E)}{dE}}{\beta(E)}.
\label{I-L-take-7}
\end{eqnarray}
Inserting formula \eqref{tra} for $\beta(E)$ into the last expression, we arrive at the desired relation \eqref{dPhi} between $\Delta\rho(E)$ and $\Phi(E)$.
Let us stress that this relation, as proven above, holds exactly, without any approximations and particularly regardless to the value of the size parameter $\varkappa$.

\section{Complex extension of classical dynamics in 1D} 
\label{AppB}

Here we sketch some features of the complex-time 1D dynamics and show that it allows for trivial tunneling trajectories exemplified in Fig.\,\ref{traj}. 
Following Refs.\,\cite{Deu10,Deu13}, we consider a classical system with ${f=1}$ in a complex phase space with coordinate ${X=\R{X}+i\I{X}}$ and momentum ${P=\R{P}+i\I{P}}$.
We use a shorthand notation ${{\rm Re}A\equiv\R{A}}$ and ${{\rm Im}A\equiv\I{A}}$ for the real and imaginary parts of a general quantity~$A$.
The complex Hamiltonian reads
\begin{eqnarray}
H(X,P)&=&\frac{P^2}{2m}+V(X)
\label{Hamacek}\\
&=&\underbrace{\left(\frac{\R{P}^2\!-\!\I{P}^2}{2m}\!+\!\R{V}(X)\right)}_{\R{H}(X,P)}
+i\underbrace{\left(\frac{\R{P}\I{P}}{m}\!+\!\I{V}(X)\right)}_{\I{H}(X,P)}.
\nonumber
\end{eqnarray}
The time $T$ is also taken complex, but we assume that it varies along a certain predetermined continuous path ${T(s)=\R{T}(s)+i\I{T}(s)}$ in the complex plane, with ${s\in{\mathbb R}}$ denoting a variable that continuously maps points along the path to real numbers.
We introduce the dot notation with the following meaning $\dot{A}=\frac{dA}{ds}=\frac{dA}{dT}\dot{T}$.

The complex Hamilton equations can be cast as
\begin{equation}
(\dot{X},\dot{P})=\left(\frac{\partial}{\partial P},-\frac{\partial}{\partial X}\right){\mathcal H}(X,P,s),
\label{Ha1}
\end{equation}
where we introduce a new Hamiltonian
\begin{equation}
{\mathcal H}(X,P,s)=H(X,P)\,\dot{T}(s).
\label{Ha2}
\end{equation}
This Hamiltonian in general depends explicitly on the time-parametrizing variable $s$ and therefore yields ${\dot{{\mathcal H}}\neq 0}$.
However, in this paper we study paths $T(s)$ such that $\dot{T}(s)$ is piecewise constant within some finite segments, so ${\cal H}$ is conseved along these segments.
Equations \eqref{Ha1} can be rewritten with the aid of the Cauchy-Riemann conditions for derivatives of a general differentiable function ${F(Z)\in{\mathbb C}}$ with respect to variable ${Z\in{\mathbb C}}$, namely
\begin{equation}
{\rm Re}\frac{dF}{dZ}\!=\!\frac{\partial\R{F}}{\partial\R{Z}}\!=\!\frac{\partial\I{F}}{\partial\I{Z}},\
{\rm Im}\frac{dF}{dZ}\!=\!\frac{\partial\I{F}}{\partial\R{Z}}\!=\!-\frac{\partial\R{F}}{\partial\I{Z}}.
\label{CauRie}
\end{equation}
Using the pair of relations with $\R{F}$, we obtain
\begin{eqnarray}
(\R{\dot{X}},\R{\dot{P}})&=&\left(\frac{\partial}{\partial\R{P}},-\frac{\partial}{\partial\R{X}}\right)\R{{\mathcal H}}(X,P,s),
\label{Haha1}\\
(\I{\dot{X}},-\I{\dot{P}})&=&\left(\frac{\partial}{\partial(-\I{P})},-\frac{\partial}{\partial\I{X}}\right)\R{{\mathcal H}}(X,P,s),
\label{Haha2}
\end{eqnarray}
where we identify two pairs of canonically conjugate real variables $(\R{X},\R{P})$ and $(\I{X},-\I{P})$ and associate the Hamiltonian with the real part $\R{{\mathcal H}}$ of Eq.\,\eqref{Ha2}. 
Thus the complexified 1D system can be treated as a system with ${f=2}$.
Using the pair of relations from Eq.\,\eqref{CauRie} with $\I{F}$, we obtain an alternative (equivalent) set of dynamical equations written in terms of the imaginary part $\I{{\mathcal H}}$ of the Hamiltonian \eqref{Ha2}, namely:
\begin{eqnarray}
({\R{\dot{X}},\R{\dot{P}}})&=&\left(\frac{\partial}{\partial\I{P}},-\frac{\partial}{\partial\I{X}}\right)\I{{\mathcal H}}(X,P,s),
\label{Ahah1}\\
({\I{\dot{X}},\I{\dot{P}}})&=&\left(\frac{\partial}{\partial\R{P}},-\frac{\partial}{\partial\R{X}}\right)\I{{\mathcal H}}(X,P,s).
\label{Ahah2}
\end{eqnarray}

As we see, the dynamics in the extended phase space of variables $(\R{X},\I{X},\R{P},-\I{P})$ depends on the selected path $T(s)$  in the complex plane of the time variable.
This path is not determined dynamically, but must be chosen {\it a priori}. 
Here, in accord with Refs.\,\cite{Deu10,Deu13}, we choose a path for which the coordinate $X(s)$ remains real during the whole motion if it starts from real initial conditions, so we can write ${X(s)=\R{X}(s)\equiv x(s)}$.
We initiate the particle at ${s=0}$ with real coordinate ${x(0)<a}$ (on the left from the interaction domain) and real momentum ${P(0)=\R{P}(0)>0}$ (pointing towards the interaction domain). 
For tunneling trajectories, the continuous path $T(s)$ looks like a descending staircase in the complex plane, the breaks being associated with transitions between the allowed and forbidden regions.
In particular, denoting ${V\equiv V\bigl(x(s)\bigr)}$, we prescribe
\begin{equation}
(\R{\dot{T}},\I{\dot{T}})
\!=\!\left\{\begin{array}{ll}
(1,0) & {\rm for\ }E\!>\!V{\ \&\ }\{E\!=\!V,\dot{V}\!<\!0\},\\
(0,-1) & {\rm for\ }E\!<\!V{\ \&\ }\{E\!=\!V,\dot{V}\!>\!0\}.
\end{array}
\right.
\label{Tdot}
\end{equation}
So while the particle is in the allowed region, the time behaves as usual, running forward along the real axis, and we have ${(\R{{\cal H}},\I{{\cal H}})=(\R{H},\I{H})}$. 
When the particle enters the forbidden regions, the time path breaks and continues running down along the negative imaginary axis, yielding ${(\R{{\cal H}},\I{{\cal H}})=(\I{H},-\R{H})}$, which lasts until the return to the allowed region on the other side of the barrier.

Trajectories in the phase space $(\R{X},\I{X},\R{P},-\I{P})$ for the time path $T(s)$ from Eq.\,\eqref{Tdot} can be calculated from Eqs.\,\eqref{Haha1}--\eqref{Haha2} with the Hamiltonian $\R{{\cal H}}$ equal to $\R{H}$ or $\I{H}$ for time running along the real or imaginary axis, respectively.
Alternatively, we can use Eqs.\,\eqref{Ahah1}--\eqref{Ahah2} with $\I{{\cal H}}$ equal to $\I{H}$ or $-\R{H}$.
The forms of Hamiltonians $\R{H}$ and $\I{H}$ follow from  Eq.\,\eqref{Hamacek}.
We point out that everywhere along the real coordinate axis ${\R{X}\equiv x}$ the potential $V(x)$ is real, so from Eq.\,\eqref{CauRie} we obtain ${\frac{\partial}{\partial\R{X}}\I{H}=-\frac{\partial}{\partial\I{X}}\R{H}=0}$ for ${X=x+i0}$.
The use of either equation pair \eqref{Haha1}--\eqref{Haha2} or \eqref{Ahah1}--\eqref{Ahah2} on the real coordinate axis then leads to
\begin{equation} 
(\R{\dot{X}},\R{\dot{P}})\!=\!\left(\frac{\R{P}}{m},-\frac{\partial V}{\partial x}\right),\
(\I{\dot{X}},\I{\dot{P}})\!=\!\left(\frac{\I{P}}{m},0\right)
\label{alow}
\end{equation}
along the $(\R{\dot{T}},\I{\dot{T}})\!=\!(1,0)$ segments, and
\begin{equation}
(\R{\dot{X}},\R{\dot{P}})\!=\!\left(\frac{\I{P}}{m},0\right),\ 
(\I{\dot{X}},\I{\dot{P}})\!=\!\left(-\frac{\R{P}}{m},\frac{\partial V}{\partial x}\right)
\label{forb}
\end{equation}
along the $(\R{\dot{T}},\I{\dot{T}})\!=\!(0,-1)$ segments.

From Eqs.\,\eqref{alow} and \eqref{forb} we first check that for a particle initiated with ${(\I{X},\I{P})=(0,0)}$ in the allowed region, the coordinate indeed remains real along the whole path, so ${\I{X}=0}$.
In the allowed region we have also ${\I{P}=0}$ and the only active variables in Eq.\,\eqref{alow} are $\R{X}$ and $\R{P}$. 
So we obtain the standard real solution of the particle motion at energy $E$.
However, as the particle reaches the classical turning point $x_{\rm T}$ with ${\R{P}\to 0_+}$ [we assume a generic turning point with ${\frac{d}{dx}V(x)|_{x_{\rm T}}>0}$], the motion does not turn back but continues to the forbidden region under the rule of Eq.\,\eqref{forb}.
While the real momentum is kept on its turning-point value ${\R{P}=0}$, the active variables become $\R{X}$ and $\I{P}$.
The dynamical equations for these variables are the same as those for $(\R{X},\R{P})$ in the allowed region, but with an inverted potential $-V(x)$.
The piecewise-conserved energies ${(\R{{\cal H}},\I{{\cal H}})}$, which were equal to $(E,0)$ in the allowed region, take values $(0,-E)$ in the forbidden region.
When the particle reaches the exit point $x'_{\rm T}$ from the forbidden region [a turning point of the inverted potential, generically satisfying ${\frac{d}{dx}V(x)|_{x'_{\rm T}}<0}$], the situation is repeated in a reversed order, the rule being returned to the dynamical equations \eqref{alow}.
For multibarrier potentials the same scenario is repeated until the particle escapes from the interaction region. 

We conclude by noting that although the above-described type of tunneling trajectories (cf.\,Fig.\,\ref{traj}) represents only one possibility out of an infinite set of various complex solutions, it seems (based on the results presented in the main text) to be really essential for the semiclassical description of the tunneling processes.

\end{document}